\documentclass[aps,prl,twocolumn,showpacs,10pt,superscriptaddress,preprintnumbers,nofootinbib]{revtex4-1}
\usepackage{epsfig,amssymb,amsmath,psfrag,epstopdf,color}
\pdfoutput=1
\usepackage{graphicx}
\usepackage{subcaption}
\usepackage{ragged2e}
\DeclareCaptionJustification{justified}{\justifying}
 \usepackage[normalem]{ulem}
\usepackage{paralist}
\usepackage{hyperref}
\usepackage[size=tiny]{todonotes}
\usepackage{xspace}

\def\beq{\begin{equation}}
\def\eeq{\end{equation}}
\def\bsp#1\esp{\begin{split}#1\end{split}}
\newcommand{\be}{\begin{equation}}
\newcommand{\ee}{\end{equation}}
\newcommand{\bea}{\begin{eqnarray}}
\newcommand{\eea}{\end{eqnarray}}

\def\cE{\mathcal{E}}

\newcommand{\PYTHIA}{\textsc{Pythia}\xspace}
\newcommand{\HERWIG}{\textsf{Herwig}\xspace}

\begin{document}

\title{Minimizing Selection Bias in Inclusive Jets in Heavy-Ion Collisions with \\ Energy Correlators}

\author{Carlota Andres}
\affiliation{Center for Theoretical Physics, Massachusetts Institute of Technology, Cambridge, MA 02139, USA}
\affiliation{Laborat\'orio de Instrumenta\c{c}\"ao e F\'isica Experimental de Part\'iculas (LIP), Av.~Prof.~Gama Pinto, 2, 1649-003 Lisbon, Portugal}

\author{Jack Holguin}
\affiliation{Consortium for Fundamental Physics, School of Physics \& Astronomy, University of Manchester, Manchester M13 9PL, United Kingdom}

\author{Raghav Kunnawalkam Elayavalli}
\affiliation{Department of Physics and Astronomy, Vanderbilt University, Nashville, TN}

\author{Jussi Viinikainen}
\affiliation{Department of Physics and Astronomy, Vanderbilt University, Nashville, TN}

\preprint{MIT-CTP/5754}

\begin{abstract}
The first-ever measurement of energy correlators within inclusive jets produced in heavy-ion collisions, revealed by the CMS Collaboration, shows a clear enhancement at large angles relative to  the proton-proton (p-p) baseline. However, interpreting this enhancement is complicated due to selection bias from energy loss, which also distorts the energy correlator heavy-ion to p-p ratio in the hadronization region, hindering our understanding of parton/hadron dynamics in a colored medium. In this \emph{Letter}, we introduce a new ratio of energy correlator observables that removes the leading effects of selection bias from the two-point energy correlator spectrum (E2C). \PYTHIA and \HERWIG simulations show that the impact of selection bias in the E2C is reduced by an order of magnitude, while sensitivity to any other medium modifications is retained. This quantity can be obtained directly from the experimental measurements presented by CMS, as illustrated in the accompanying note~\cite{Andres:2024pyz}. 
\end{abstract}

\maketitle


\emph{Introduction.}--- The first-ever measurement of energy correlators in heavy-ion collisions, recently unveiled by the CMS Collaboration~\cite{talkEEC,CMS-PAS-HIN-23-004}, unlocks a new frontier in the study of the quark-gluon plasma (QGP)~\cite{Busza:2018rrf,Cao:2020wlm,Arslandok:2023utm,CMS:2024krd,ALICE:2022wpn,Apolinario:2022vzg, Cunqueiro:2021wls,Connors:2017ptx}, paving the way for re-imagining jet substructure through correlation functions of energy flow operators~\cite{Basham:1977iq,Basham:1978bw,Basham:1978zq,Basham:1979gh,Hofman:2008ar}. Several distinctive features of energy correlators were crucial to this landmark achievement. Notably, energy correlators  can be both theoretically computed and experimentally measured on charged particles~\cite{Chang:2013rca,Chang:2013iba,Li:2021zcf,Chen:2022muj,Chen:2022pdu,Jaarsma:2022kdd,Jaarsma:2023ell,Barata:2024nqo}. Moreover, the inherently statistical nature of both the observable and the uncorrelated heavy-ion background enabled the application of a sophisticated, data-driven subtraction method, which estimates the residual background by considering cross-correlations between events with and without reconstructed jets~\cite{talkEEC,CMS-PAS-HIN-23-004}. These advances collectively enabled a direct comparison between the experimental measurement  and  several theoretical predictions~\cite{Andres:2024ksi, Andres:2022ovj, Andres:2023xwr,Yang:2023dwc,Bossi:2024qho}, revealing a particularly good qualitative agreement with the semi-analytical calculations presented in~\cite{Andres:2024ksi}.

In the heavy-ion two-point energy correlator  spectrum  (E2C) measured by CMS in inclusive jet events, two notable effects are observed. First, in the large angular regime\footnote{ We note that since our focus is on the collinear limit of the energy correlators measured within a jet, where the angle is always smaller than 1,  ``large angle''  refers to the regime where the angle is much greater than $\Lambda_{\rm QCD}/p_{T}$, $p_T$ being the reconstructed jet transverse momentum.}, an enhancement of the E2C with energy weight $n=1$ is seen in central to semi-peripheral lead-lead (Pb-Pb) collisions relative to proton-proton (p-p) collisions. While this large-angle enhancement, attributed to relatively soft particles, had been predicted by various theoretical calculations~\cite{Andres:2022ovj,Andres:2023xwr, Andres:2024ksi,Bossi:2024qho,Yang:2023dwc}, its precise nature is not yet fully understood. It could be driven by different factors, including medium-induced splittings and transverse momentum broadening~\cite{Andres:2022ovj,Andres:2023xwr,Andres:2024ksi,Barata:2023bhh}, medium response~\cite{Bossi:2024qho,Yang:2023dwc} or a combination of these effects. Disentangling the impact of these varying phenomena motivates further studies, including higher-order energy correlators~\cite{Bossi:2024qho}.

Second, the E2C of inclusive jets in Pb-Pb collisions is shifted to smaller angles compared to that of inclusive jets in p-p collisions at the same reconstructed $p_T$. This shift is a result of selection bias:  heavy-ion jets lose energy in the QGP,  so comparing inclusive jet samples in p-p and heavy-ion (A-A) collisions at the same reconstructed $p_T$
corresponds to different jet populations produced at different hard scales. In particular, this energy loss distorts the view of the hadronization transition in the E2C spectrum and complicates the extraction of scaling laws from the large-angle enhancement. 
Since the hadronization transition has been successfully measured across a broad range of jet $p_T$'s in p-p collisions from LHC~\cite{CMS:2024mlf,talk2,talk4,talk6} to RHIC~\cite{Tamis:2023guc}, it is highly desirable to be able study any potential medium modifications to hadronization in heavy-ion collisions without the confounding effects of the predicted and observed selection bias~\cite{Andres:2024ksi,Yang:2023dwc,talkEEC,CMS-PAS-HIN-23-004}. In addition, removing energy loss effects is instrumental for accurately extracting  any  QGP-induced modifications to jet substructure.

Selection bias driven by energy loss has long been a major obstacle in disentangling jet substructure modifications in heavy-ion collisions. As a result, extensive literature has focused on studying or mitigating its effects~\cite{Baier:2001yt,Zhang:2009rn,dEnterria:2009xfs,Renk:2012ve,Milhano:2015mng,Spousta:2015fca,Rajagopal:2016uip,Casalderrey-Solana:2015vaa,Casalderrey-Solana:2016jvj,Caucal:2018dla, Casalderrey-Solana:2018wrw, Brewer:2018dfs,Casalderrey-Solana:2019ubu,Caucal:2020xad,Du:2020pmp,Du:2021pqa,Apolinario:2021olp,Takacs:2021bpv,Brewer:2021hmh}. A recent effective approach involves using $\gamma/Z~+~$jet events, where selecting jets based on the unmodified $p_T^B$ of the recoiling boson and on the momentum imbalance $x_J=p_T/p_T^B$~\cite{Brewer:2021hmh,CMS:2024zjn,CMS:2017eqd,CMS:2017ehl,ATLAS:2018dgb,CMS:2018jco,ATLAS:2020wmg} has been shown to significantly reduce this bias in substructure observables~\cite{Brewer:2021hmh,CMS:2024zjn} \footnote{Indeed, the E2C was originally proposed to be measured in heavy-ion collisions on $\gamma/Z~+~$jet  samples, where energy loss effects are subleading~\cite{Andres:2022ovj,Andres:2023xwr}.}. Furthermore, tagging a jet with a recoiling vector boson could enable the measurement of the E2C across the full angular regime, extending beyond the collinear limit. However, $\gamma/Z~+~$jet processes are much rarer than inclusive jet events.  The significantly higher statistics available for inclusive jets, combined with their greater sensitivity to gluon-initiated jets, make it essential to continue to use inclusive jet events to probe the QGP. Achieving this requires a thorough approach to disentangle physical effects from selection bias.

In this \emph{Letter}, we present a novel procedure to mitigate selection bias that capitalizes on the unique properties of energy correlators. By carefully constructing double ratios involving these observables, it becomes possible to directly cancel the leading selection bias effects in inclusive heavy-ion jet samples. Our method draws parallels with the $\alpha_{\rm s}$-extraction strategy used in p-p collisions, where specific correlator ratios eliminate classical dynamics from the spectrum~\cite{Chen:2020vvp,Komiske:2022enw,Lee:2022ige,CMS:2024mlf}. In the following, we derive the \emph{unbiasing function} that corrects for energy loss effects in  inclusive jet E2C measurements and demonstrate its effectiveness with simulated events from \PYTHIA and \HERWIG where energy loss is modeled as a shift in the jet $p_{T}$ spectrum. The supplemental note~\cite{Andres:2024pyz} further validates this method by applying it to experimentally measured E2C distributions from CMS~\cite{talkEEC,CMS-PAS-HIN-23-004}.

\emph{Cancelling Selection Bias.}--- We define the $N$-point projected energy correlator (ENC)~\cite{Chen:2020vvp,Komiske:2022enw} as
\begin{align}
f_{\rm ENC}(R_{L}) \equiv \mathcal{N} \int &\prod^N_{i=1}{\rm d}^2\vec{n}_i\, \langle \,   \cE(\vec{n}_1) \cdots \cE(\vec{n}_N)  \,\rangle \nonumber \\
& \times \delta\left(R_{L} - \Delta \hat{R}_L(\vec{n}_1, \cdots ,\vec{n}_N) \right)\,,
\label{eq:ENC}
\end{align}
where $\langle \cdots \rangle$ denotes the expectation value over the state in which the correlator is evaluated, $\hat{R}_{L}$ is an operator that selects the largest boost invariant angle among the measurement directions ($\vec{n}$) of the $N$-tuple of energy flow operators~\cite{Sveshnikov:1995vi,Tkachov:1995kk,Korchemsky:1999kt,Bauer:2008dt,Hofman:2008ar,Belitsky:2013xxa,Belitsky:2013bja,Kravchuk:2018htv}, and $\mathcal{N}$ is a normalization. In terms of inclusive cross-sections, the ENC is expressed as 
\begin{align}
f_{\rm ENC}(R_{L}) = \mathcal{N} \int \frac{{\rm d} \sigma_{1,\cdots, N}}{{\rm d} \Delta \hat{R}_L } \,E_{1}\cdots E_{N} \, \delta\left(R_{L} - \Delta \hat{R}_L\right) \,,
\end{align}
where ${\rm d} \sigma_{1,\cdots, N}$ is the cross-section to produce $N$ hadrons and $E_{i}$ represents their respective energies. For convenience, we assume that $\mathcal{N}$ is chosen so that $f_{\rm ENC}$ has amplitude $1$ at the hadronization peak $R_{\rm peak}$, i.e., $f_{\rm ENC}(R_{\rm peak})=1$. This normalization simplifies the algebraic derivation of the unbiasing function, but our final result applies to ENCs with any normalization.

Additionally, we define a generalized cumulative distribution function of the ENC as
\begin{align}
F_{\rm ENC}(R_{L},p) \equiv \int^{R_{L}}_{0} {\rm d} R \,\left(f_{\rm ENC} (R) \right)^{p}\,.
\label{eq:cumulant}
\end{align}

In p-p collisions, features in the ENC spectrum are governed by a single dimensionful quantity $p^{\rm hard}_T R_L$, where $p^{\rm hard}_T$ represents the initial transverse momentum of the parton from the hard process that results in the jet. This distinctive and powerful characteristic of p-p ENCs  emerges from their factorization in the collinear limit onto a single jet-function $J(R_{L}, \ln(R_{L} p^{\rm hard}_T / \mu) )$ \cite{Dixon:2019uzg} and has been experimentally verified over two orders of magnitude in $p^{\rm hard}_T$~\cite{Tamis:2023guc,talk1, talk2,talk4,talk6,CMS:2024mlf}. Therefore, leaving additional effects from the A-A environment for later consideration, selection bias due to energy loss modifies the A-A ENC, with our chosen normalization, as follows:
\begin{align}
    f^{\rm AA}_{\rm ENC}(R_{L}) = \int \mathrm{d}\varepsilon ~ \bar{p}(\varepsilon) f^{\rm pp}_{\rm ENC}\left(R_{L}\left(1 + \frac{\varepsilon \, P(R_L)}{p_{T}} \right) \right)\,, \label{eq:ENCshifted}
\end{align}
where $\varepsilon \ll p_{T}$ represents the small energy loss of the heavy-ion jet with reconstructed transverse momentum $p_{T}$, such that $p_{T} + \varepsilon$ corresponds to the initial hard scale $p^{\rm hard}_{T}$. The term $\bar{p}(\varepsilon)$ denotes the probability of an energy loss $\varepsilon$ occurring in a measured heavy-ion jet. The factor $P(R_{L})\sim 1$ accounts for the angular dependence of energy loss, introducing a bias toward narrower heavy-ion jets.  Since $P(R_{L})$ varies modestly with $R_{L}$~\cite{Baier:2001yt,Salgado:2003gb, Mehtar-Tani:2022zwf,Barata:2023bhh}, we assume for power counting in expansions that $P(R_{L}) = 1 + \mathcal{O}(\varepsilon/p_{T})g(R_L)$, where $g(R_L)\sim 1$.

Using~\eqref{eq:ENCshifted}, the modification of the A-A generalized cumulative distribution function  due to energy loss can be written as: 
\begin{align}
    & F^{\rm AA}_{\rm ENC}(R_{L},p) =  \int \mathrm{d}\varepsilon \, \bar{p}(\varepsilon) \Bigg[ \left(1 + \frac{\varepsilon \,P(R_L)}{p_{T}} \right)^{-1} \nonumber \\ & ~~~~\times F^{\rm pp}_{\rm ENC}\left(R_{L}\left(1 + \frac{\varepsilon \,P(R_L)}{p_{T}} \right) , p \right) + \mathcal{O}\left(\frac{\varepsilon^2}{p_{T}^2}\right) \Bigg]\,.
\end{align}
Hence, the A-A to p-p ratios for the ENC and its cumulative distribution function, are respectively given by:
\begin{align}
    \frac{f^{\rm AA}_{\rm ENC}(R_{L})}{f^{\rm pp}_{\rm ENC}(R_{L})} = 1 + \frac{\bar{\varepsilon} \, P(R_L)}{p_{T}} \frac{{\rm d} \ln f^{\rm pp}_{\rm ENC} (R_L) }{ {\rm d} \ln R_{L}}\, + \mathcal{O}\left(\frac{\overline{\varepsilon^2}}{p_{T}^2}\right)\,,
    \label{eq:enc_ratio0}
\end{align}
where $\bar{\varepsilon}=\int \mathrm{d} \varepsilon ~ \bar{p}(\varepsilon) \varepsilon$ and $\overline{\varepsilon^2}=\int \mathrm{d} \varepsilon ~ \bar{p}(\varepsilon) \varepsilon^2$,  and 
\begin{align}
    &\frac{F^{\rm AA}_{\rm ENC}(R_{L},p)}{F^{\rm pp}_{\rm ENC}(R_{L},p)} = \nonumber \\
    &1 + \frac{\bar{\varepsilon} \,P(R_L)}{p_{T}} \left(\frac{{\rm d} \ln F^{\rm pp}_{\rm ENC} (R_{L},p) }{ {\rm d} \ln R_{L}} - 1\right)+ \mathcal{O}\left(\frac{\overline{\varepsilon^2}}{p_{T}^2}\right)\,.
    \label{eq:cu_ratio0}
\end{align}
We observe that the leading correction due to energy loss has the same basic form in Eqs.~\eqref{eq:enc_ratio0}~and~\eqref{eq:cu_ratio0}. Therefore, if the derivative in~\eqref{eq:enc_ratio0} can be arranged to cancel the bracket in~\eqref{eq:cu_ratio0}, energy loss can be effectively removed from~\eqref{eq:enc_ratio0} by dividing it by a distribution based on~\eqref{eq:cu_ratio0}, without requiring complete knowledge of $\varepsilon P(R_L)$.

To achieve this, we first examine the derivative terms in Eqs.~\eqref{eq:enc_ratio0}~and~\eqref{eq:cu_ratio0}, focusing, from now on, on the specific case of the two-point energy correlator, $f_{\rm E2C}$, recently measured in both  p-p and heavy-ion collisions~\cite{talk2, Tamis:2023guc, talk4, talk6, CMS:2024mlf,talkEEC,CMS-PAS-HIN-23-004}. In the perturbative regime where $R_{L} \gg \Lambda_{\rm QCD}/p_{T}$, the leading-order scaling of the p-p $f_{\rm E2C}$ is $\left.f_{\rm E2C}^{\rm {pp}}\right|_{\rm pert.} \sim R_L^{-1+\mathcal{O}(\alpha_{\rm s})}$~\cite{Hofman:2008ar, Kologlu:2019mfz,Dixon:2019uzg}. Hence, for  $p > 1$  the following relations hold
\begin{align}
    &\left. \frac{{\rm d} \ln f^{\rm pp}_{\rm E2C}(R_L) }{ {\rm d} \ln R_{L}} \right|_{\rm pert.} = -1 + \mathcal{O} \left(\alpha_{\rm s}, \frac{\Lambda_{\rm QCD}}{R_{L}p_T}\right) \, ,  \label{eq:8}\\ 
    &\left. \frac{{\rm d} \ln F^{\rm pp}_{\rm E2C} (R_{L},p)}{ {\rm d} \ln R_{L}}\right|_{\rm pert.} = 0 + \mathcal{O} \left(\alpha_{\rm s},\left( \frac{\Lambda_{\rm QCD}}{R_{L}p_T}\right)^{p-1}\right)\, , \nonumber
\end{align}
where we have used $\Lambda_{\rm QCD}/p_T$ as an infrared cut-off for $R_{L}$. To study the region where $R_{L} \lesssim \Lambda_{\rm QCD}/p_T$, non-perturbative (NP) information is required. At very small angles, $R_{L} \ll \Lambda_{\rm QCD}/p_T$, the p-p E2C exhibits the scaling behavior of a free hadron gas, i.e., $\left. f^{\rm pp}_{\rm E2C} \right|_{\rm f.h}\sim R_{L}$~\cite{Komiske:2022enw,talk2,Tamis:2023guc,CMS:2024mlf}. In this regime
\begin{align}
    &\left. \frac{{\rm d} \ln f^{\rm pp}_{\rm E2C}(R_L) }{ {\rm d} \ln R_{L}} \right|_{\rm f.h.} = 1 \, , ~~~ \left. \frac{{\rm d} \ln F^{\rm pp}_{\rm E2C} (R_{L},p)}{ {\rm d} \ln R_{L}}\right|_{\rm f.h.} = p+1 \, . \label{eq:fhgas}
\end{align}
Interpolating between these two limits, we obtain that
\begin{align}
    \frac{{\rm d} \ln f^{\rm pp}_{\rm E2C}(R_L) }{ {\rm d} \ln R_{L}} \approx \frac{2}{p+1}\frac{{\rm d} \ln F^{\rm pp}_{\rm E2C} (R_{L},p)}{ {\rm d} \ln R_{L}} - 1 . 
    \label{eq:approx}
\end{align}

To assess the performance of Eq.~\eqref{eq:approx} across the entire range $R_L<1$, and particularly around the NP transition where $R_L \sim \Lambda_{\rm QCD}/p_T$, a deeper understanding of the NP effects is necessary~\cite{Lee:2022ige,Hofman:2008ar,Schindler:2023cww,Lee:2024esz,Chen:2024nyc}. To this end, we computed both sides of Eq.~\eqref{eq:approx} using the experimental data on the E2C spectrum~\cite{CMS-PAS-HIN-23-004}.  Our analysis shows that the equality holds well over the full experimentally measured range, $0<R_{L}<0.4$, for both $p=2$ and $p=3$~\cite{Andres:2024pyz}. The accuracy achieved was $10\%$ or better for perturbatively large values of $R_L$
and between $10\%$ and $20\%$ in the non-perturbative region. Relevant figures are provided in the supplementary note~\cite{Andres:2024pyz}.

We can now replace~\eqref{eq:approx} into \eqref{eq:cu_ratio0} and rearrange it to obtain 
\begin{align}
    &\left(\frac{F^{\rm AA}_{\rm E2C}(R_{L},p)}{F^{\rm pp}_{\rm E2C}(R_{L},p)}\right)^{\frac{2}{p+1}} - \frac{\bar{\varepsilon} \,P(R_L)}{p_{T}}  \frac{p-1}{p+1}   \nonumber \\
    \,
    &~~~ \approx 1 + \frac{\bar{\varepsilon} \,P(R_L)}{p_{T}} \frac{{\rm d} \ln f^{\rm pp}_{\rm E2C} (R_{L}) }{ {\rm d} \ln R_{L}}\,,
    \label{eq:cu_ratio}
\end{align}
where the right-hand side corresponds to that of Eq.~\eqref{eq:enc_ratio0}. Thus, the left-hand side of \eqref{eq:cu_ratio} provides the necessary means to cancel the leading-order effects of energy loss in the A-A E2C. The only unknown factor, $\bar{\varepsilon} \,P(R_L) /p_{T}$, can  be determined in a data driven way by identifying the $R_{L}$ bin where
\begin{align}
  \frac{{\rm d} \ln f^{\rm pp}_{\rm E2C} (R_{L}) }{ {\rm d} \ln R_{L}}\Bigg|_{R_{L}=R_{\rm peak}} =0\,,  
\end{align}
which corresponds to the position of the hadronization peak $R_{\rm peak}$ in the p-p E2C spectrum. Then, at $R_L=R_{\rm peak}$, Eq.~\eqref{eq:cu_ratio} can be solved to find  the dimensionless quantity $E_{\rm peak}$
\begin{align}
     E_{\rm peak} \equiv \frac{\bar{\varepsilon} \,P(R_{\rm peak})}{p_{T}} \approx  
     \frac{p+1 }{p-1}   \left[\left(\frac{F^{\rm AA}_{\rm ENC}(R_{\rm peak},p)}{F^{\rm pp}_{\rm ENC}(R_{\rm peak},p)}\right)^{\frac{2}{p+1}} - 1 \right]\,.
    \label{eq:unknownfactor}
\end{align}
We can now introduce the \emph{unbiasing function} $C_p$ which cancels the $R_L$-dependent effects of energy loss in the heavy-ion E2C, up to corrections of  order $\overline{\varepsilon^2}/p_T^2$:
\begin{align}
    C_p(R_{L})\equiv \left(\frac{F^{\rm AA}_{\rm ENC}(R_{L},p)}{F^{\rm pp}_{\rm ENC}(R_{L},p)}\right)^{\frac{2}{p+1}} - E_{\rm peak} ~  \frac{p-1}{p+1}  \, ,
    \label{eq:cu_ratio2}
\end{align}
where  $E_{\rm peak}$ is evaluated using \eqref{eq:unknownfactor}, since $P(R_{\rm peak})-  P(R_L) = \mathcal{O}(\varepsilon/p_T)$ for $0 < R_L < 1$. We have observed that experimental error bands on $C_p$ are smallest for lower values of $p$, making $p=2$ the optimal choice. Hence, our proposed correlator-based observable, which minimizes selection bias effects, is given by
\begin{align}
    {\rm E2C}/C_{2} \equiv f^{ \rm AA}_{\rm E2C}(R_{L}) \left/ C_{2}(R_{L})\,.\right.
    \label{eq:unbiased_spec}
\end{align}
Additionally, we define its ratio against the p-p E2C spectrum $f^{{\rm pp}}_{\rm E2C}$ as
\begin{align}
    {\rm rE2C}/C_{2} \equiv \left.\frac{f^{ \rm AA}_{\rm E2C}(R_{L})}{f^{ \rm pp}_{\rm E2C}(R_{L})} \right/ C_{2}(R_{L})\, .
     \label{eq:unbiased_ratio}
\end{align} 
We highlight that $C_2$ should be obtained using our prescribed normalization for the E2C spectrum,  but, once determined, it can be applied to the E2C with any normalization.   Equivalent unbiasing functions can readily be derived for other single-variable energy-correlator observables, such as the ENC or $\left<\mathcal{E}^2 \mathcal{E}^2\right>$, by appropriately modifying Eqs.~\eqref{eq:8}~and~\eqref{eq:fhgas}.


\emph{General medium modifications.}--- In the following, we aim to show that while selection bias due to energy loss is effectively canceled in the  ${\rm E2C}/C_2$ and ${\rm rE2C}/C_2$
distributions, other QGP-driven modifications are not.

We start by expressing the general A-A ENC as
\begin{align}
    f^{{\rm AA}}_{\rm ENC}(R_{L}) = \int \mathrm{d}\varepsilon ~\,\bar{p}(\varepsilon) \, f^{{\rm pp}}_{\rm ENC}(R_{L}')\,(1 + \delta f^{{\rm AA}}_{\rm ENC}(R_{L}'))\,,
\end{align}
where $R_{L}'\equiv R_{L}\left(1 + \varepsilon \, P(R_L)/p_{T} \right)$. 
We assume that the heavy-ion ENC is only moderately modified with respect to the p-p ENC, such that $\delta f^{{\rm AA}}_{\rm ENC}(R_{L'}) \lesssim 1$ for $0<R_L'<1$. Additionally, we consider that, apart from energy loss effects, the QGP is not expected to significantly modify late-time free-hadron correlations. Thus, QGP-driven modifications of $f^{{\rm AA}}_{\rm ENC}$ are assumed to be restricted to a region $R_{L}>R_{\rm mod}$ with $R_{\rm mod} \gtrsim \Lambda_{\rm QCD}/p_{T}$, so that $\delta f^{{\rm AA}}_{\rm ENC}(R_{L}< R_{\rm mod}) \approx 0$. This is further supported by recent theoretical calculations of the heavy-ion E2C within high-energy ($p_T>100~$GeV) jets, which find that, aside from energy loss, QGP-driven modifications to the E2C become significant only for  $R_{L} \gg \Lambda_{\rm QCD}/p_{T}$~\cite{Andres:2024ksi,Yang:2023dwc,Barata:2023bhh,Bossi:2024qho}. Within these assumptions, we obtain that
\begin{align}
   \Delta&=\int^{R_L}_{0} {\rm d} R \,  \delta f^{{\rm AA}}_{\rm ENC}(R)\, \frac{ \big( f^{{\rm pp}}_{\rm ENC}(R) \big)^p }{F^{{\rm pp}}_{\rm ENC}(R_{L},p)} \nonumber \\
    &\approx \int^{R_{L}}_{R_{\rm mod}} {\rm d} R \, \delta f^{{\rm AA}}_{\rm ENC}(R)\, \frac{ \big( f^{{\rm pp}}_{\rm ENC}(R) \big)^p }{F^{{\rm pp}}_{\rm ENC}(R_{L},p)} \ll 1\,,
    \label{eq:smallinCum}
\end{align}
since we are integrating the product of two functions which are each maximally $\sim 1$ over a range that is significantly less than $1$.

\begin{figure}
    \centering
    \includegraphics[scale=0.43]{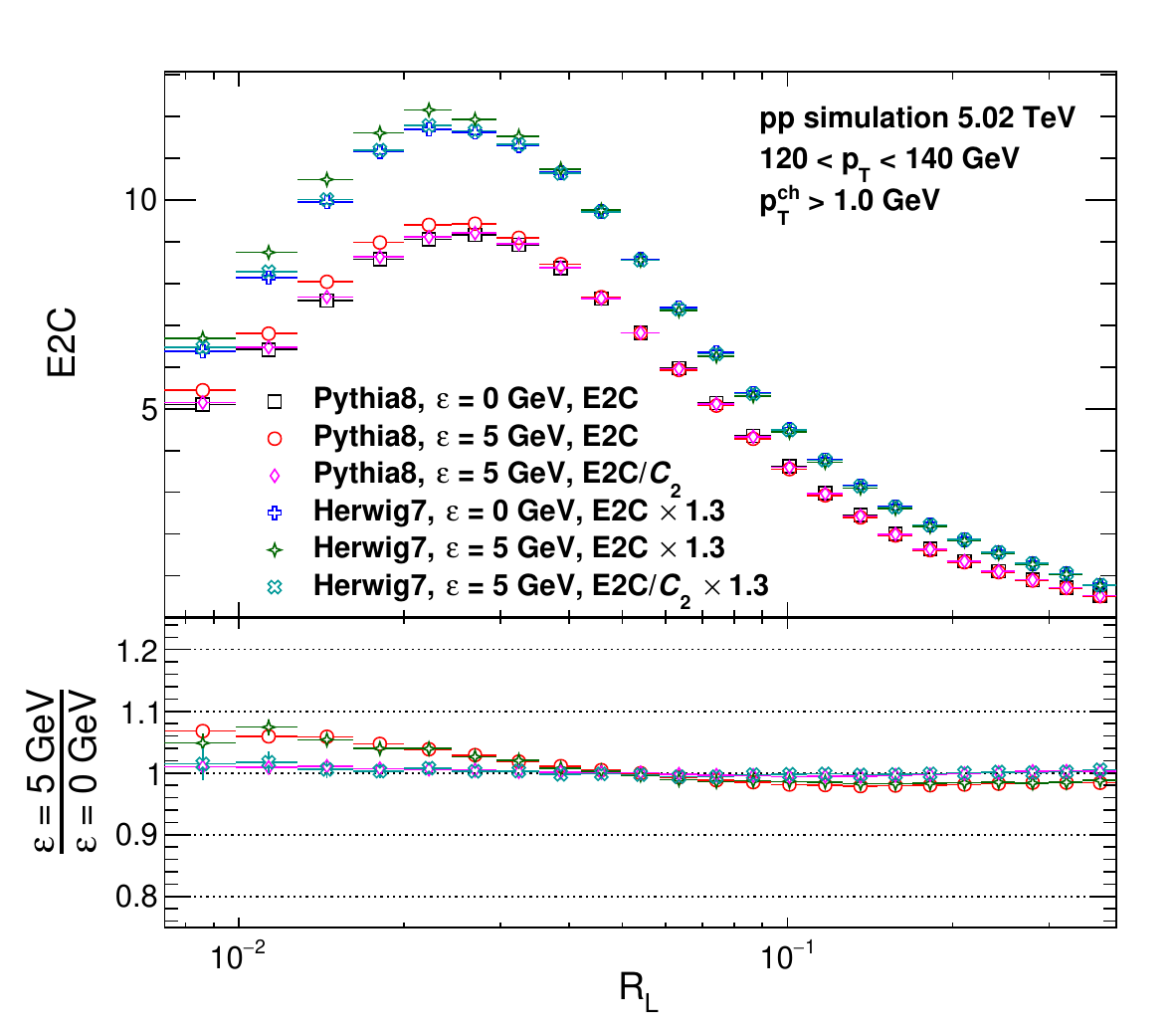}
    \caption{Top panel: E2C and E2C/$C_2$ for inclusive jets with  $120<p_T<140~$GeV (labelled $\varepsilon=0~$GeV) compared to those with $125<p_T<145~$GeV, corresponding to a constant selection bias shift of $\varepsilon=5~$GeV. Bottom panel:  E2C and E2C/$C_2$ for $125<p_T <145~$GeV jets divided by the E2C of $120<p_T<140~$GeV jets.}
    \label{fig:5GeV}
\end{figure}

Turning our attention to the proposed observable rE2C/$C_2$, we can expand it in terms of the energy loss $\varepsilon$ and Eq.~\eqref{eq:smallinCum} to find
\begin{align}
    &{\rm rE2C}/C_2 = 1 + \int \mathrm{d}\varepsilon \,\bar{p}(\varepsilon) \, \Bigg[ \delta f^{\rm AA}_{\rm E2C}(R_{L}') \,-\nonumber \\
    &   \frac{4 R_{L}}{3 R_{L}'} \int^{R'_{L}}_{R_{\rm mod}} {\rm d} R ~ \delta f^{ \rm AA}_{\rm E2C}(R) \, \frac{f^{\rm pp}_{\rm E2C}(R)^{2}}{F^{\rm pp}_{\rm E2C}(R_{L},2)} \Bigg]+ \mathcal{O}\left(\frac{\overline{\varepsilon^2}}{p_{T}^2} , \Delta^2 \right), \nonumber \\
    & =  1 + \delta f^{\rm AA}_{\rm E2C}(R_{L}) + \mathcal{O}\left(\frac{\bar{\varepsilon}}{p_{T}}  \delta f^{\rm AA}_{\rm E2C}\right)\,,
    \label{eq:itworks}
\end{align}
which demonstrates that the proposed double ratio \eqref{eq:unbiased_ratio} is highly sensitive to medium modifications of the E2C spectrum while effectively canceling the distortions caused by energy loss at leading order.  

Moreover, we note that the term in the second line of Eq.~\eqref{eq:itworks} becomes negligibly small when $R_{\rm mod} \gg \Lambda_{\rm QCD}/p_{T}$, making the observable particularly well-suited for investigating large-angle medium modifications to the E2C spectrum. When $R_{\rm mod} \sim \Lambda_{\rm QCD}/Q$ this term remains suppressed but may have a more noticeable numerical impact. Consequently, the ${\rm E2C}/C_2$ and ${\rm rE2C}/C_2$ distributions might exhibit slightly reduced sensitivity to medium modifications in the hadronization transition compared to the large-angle regime. Nonetheless, these observables still ensure that any significant changes observed in this regime relative to p-p cannot be attributed to selection bias.

\begin{figure}
    \centering
    \includegraphics[scale=0.43]{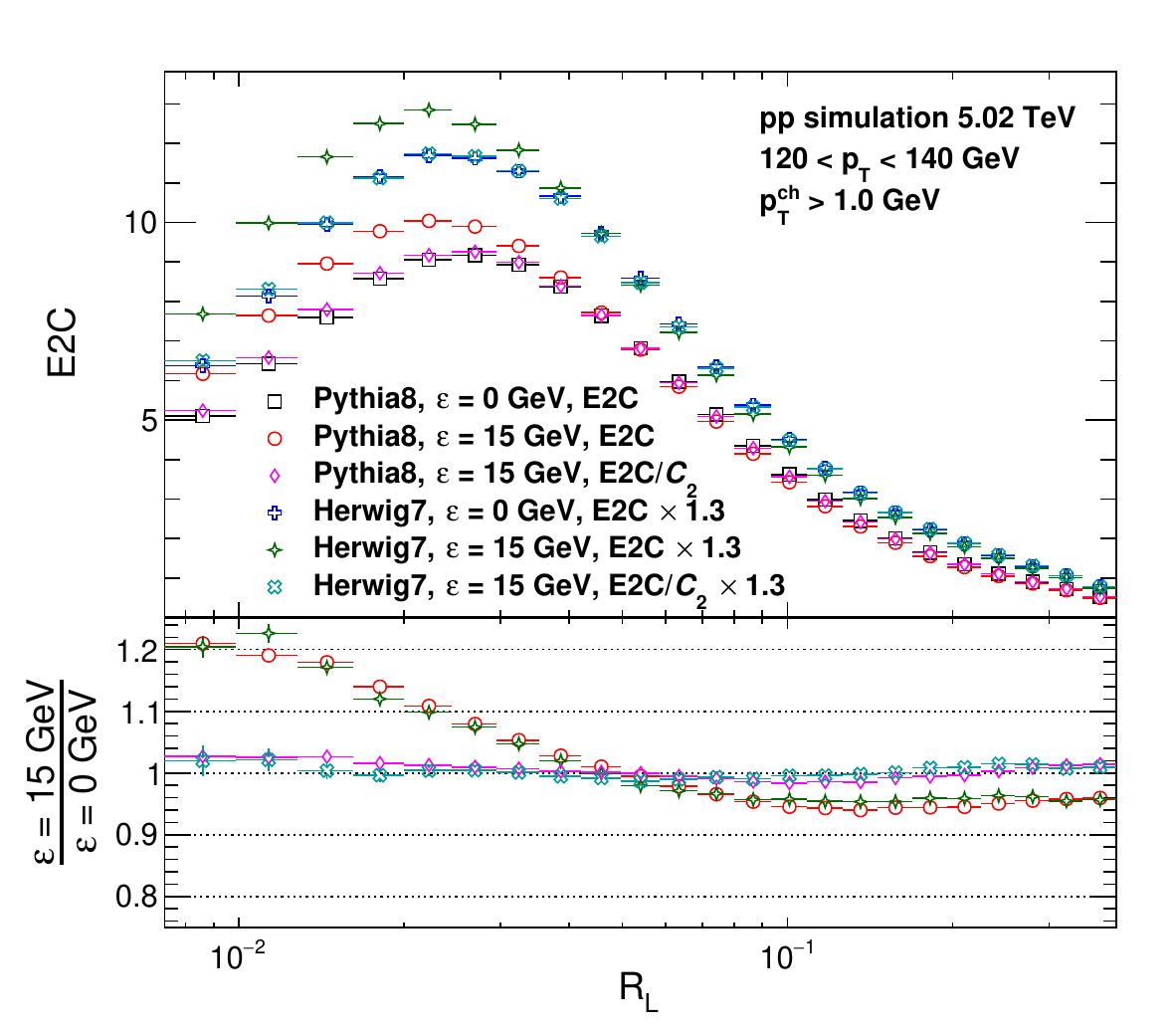}
    \caption{Top panel: E2C and E2C/$C_2$ for inclusive jets  with $120<p_T<140~$GeV (labelled $\varepsilon=0~$GeV) compared to those with $135<p_T<155~$GeV, corresponding to a constant selection bias of $\varepsilon=15~$GeV. Bottom panel:  E2C and E2C/$C_2$ for $135<p_T <155~$GeV jets divided by the E2C of $120<p_T<140~$GeV jets.}
    \label{fig:15GeV}
\end{figure}

\emph{Numerical results.}--- In Figs.~\ref{fig:5GeV} and \ref{fig:15GeV}, we evaluate the efficacy of the proposed unbiasing function $C_{2}$ using inclusive jet samples in $\sqrt{s}=5.02~$TeV p-p collisions generated with both \PYTHIA 8.230~\cite{Sjostrand:2014zea} and \HERWIG 7.2.2~\cite{Bellm:2015jjp,Bahr:2008pv} Monte Carlo event generators. For \PYTHIA, around 44 M events with tune CP5~\cite{CMS:pythia8tune} were generated, while for \HERWIG around 7 M events with tune CH3~\cite{CMS:herwig7tune} were generated. Inclusive jets were clustered from these samples using the anti-$k_T$ algorithm~\cite{Cacciari:2008gp} with a jet radius of $R=0.4$. The E2C distribution was computed for charged particles around these jets following the analysis strategy outlined in the recent CMS heavy-ion E2C measurements~\cite{talkEEC,CMS-PAS-HIN-23-004}. 

To analyze the effects of energy loss, we compare jets samples with slightly different reconstructed $p_{T}$ ranges. We treat each sample with slightly higher $p_{T}$ as an A-A sample, with its energy loss $\varepsilon$ defined as the difference in the mean $p_{T}$ between the considered samples. Crucially, this approach models the A-A system as differing from p-p only by the selection bias induced by energy loss. We use both \PYTHIA and \HERWIG to demonstrate that the cancellation of selection bias in ${\rm E2C}/C_{2}$ is achieved independently of the hadronization and parton shower modeling considered.

Fig.~\ref{fig:5GeV} illustrates our results for a selection bias of $\varepsilon=5~$GeV on a sample of $\sim 120~$GeV jets. The bottom panel shows that, in contrast with the E2C A-A/p-p ratio (red and green markers), the E2C/$C_2$ ratio (blue and pink markers) remains almost completely flat in the full $R_L$-range, indicating that the proposed unbiasing function $C_2$ effectively removes the selection bias from the E2C spectrum. Specifically, without the $C_2$ function, the E2C  exhibits features due to energy loss that are as large as approximately $10\%$ in magnitude.  When $C_{2}$ is included, these features are reduced to $\lesssim 1\%$ . This result confirms that the procedure outlined in this \emph{Letter} is highly effective in mitigating the impact of small biases. 

Fig.~\ref{fig:15GeV} shows a selection bias of $\varepsilon=15~$GeV on the same sample of jets. Without the unbiasing function $C_2$, this selection bias significantly affects the E2C, with deviations reaching up to $20\%$ (see red and green markers in the bottom panel). However, when  $C_{2}$ is applied, the effect of selection bias on the E2C is almost entirely removed at large angles, reducing it to around a $1\%$ effect (see blue and pink markers). At smaller angles, the impact of selection bias is reduced to $\sim 2\%$. 

In the supplemental material, we confirm that these results also hold for jets with higher transverse momentum, ranging between $120$ and $200~$GeV, for both a selection bias of $\varepsilon=5~$GeV and  $\varepsilon=15~$GeV.  Furthermore, in an accompanying note~\cite{Andres:2024pyz} we demonstrate that our methodology can be successfully applied to  experimental data~\cite{talkEEC,CMS-PAS-HIN-23-004}. Overall, selection bias is consistently reduced by an order of magnitude in the E2C/$C_2$ and the E2C/$C_2$ A-A/p-p ratio.

\emph{Conclusions and Outlook.}--- In this \emph{Letter} we have introduced a novel energy correlator-based observable with greatly reduced sensitivity to selection bias from energy loss in heavy-ion inclusive jet samples. Our proposed double ratio is the first jet substructure observable, which can be measured in heavy-ion inclusive jets, where energy loss does not play a leading role. We derived this result analytically and demonstrated its efficacy through event generator simulations. These simulations show that our proposed observable is affected by energy loss related shift in the jet spectrum by an order of magnitude less than the standard two-point energy correlator.

The recent unveiling of the first measurement of energy correlators in heavy-ion collisions~\cite{talkEEC,CMS-PAS-HIN-23-004} followed a growing theoretical interest in these observables~\cite{Andres:2022ovj,Andres:2023xwr, Andres:2024ksi,Andres:2023ymw,Barata:2023bhh,Barata:2023zqg,Bossi:2024qho,Yang:2023dwc,Singh:2024vwb,Holguin:2022epo,Holguin:2023bjf,Holguin:2024tkz,Liu:2022wop,Liu:2023aqb,Cao:2023oef}. To deepen our understanding of these observables, it is essential to remove selection bias. As a first step, we show in an accompanying note~\cite{Andres:2024pyz} that applying our methodology to the current available experimental E2C data on inclusive jets~\cite{talkEEC,CMS-PAS-HIN-23-004}, reveals physical modifications to two-point energy correlations attributable  to the heavy-ion environment. Additionally, our procedure could be applied to future heavy-ion $N$-projected energy correlator measurements on inclusive, di-jets and gamma/$Z$-jets samples, enabling access to different $q/g$ fractions without compounding effects of energy loss. We anticipate that our approach will set a standardized strategy for mitigating selection bias in future experimental measurements of energy correlators.

\emph{Acknowledgements.}-- We thank Fabio Dominguez, Ian Moult and Krishna Rajagopal for their valuable discussions and comments on the manuscript. We are also thankful to Hannah Bossi, Kyle Lee, Yen-Jie Lee, Cyrille Marquet, and Carlos A. Salgado for useful discussions. The work of CA was partially supported by the U.S. Department of Energy, Office of Science, Office of Nuclear Physics under grant Contract Number DE-SC0011090 and by OE Portugal, Funda\c{c}\~{a}o para a Ci\^{e}ncia e a Tecnologia (FCT), I.P., project 2024.06117.CERN. CA acknowledges the financial support by the FCT under contract  2023.07883.CEECIND. RKE would like to acknowledge funding by the U.S. Department of Energy, Office of Science, Office of Nuclear Physics under grant number DE-SC0024660. JV would like to acknowledge funding by the U.S. Department of Energy, under grant number DE-FG05-92ER40712. The authors would like to express special thanks to the Mainz Institute for Theoretical Physics (MITP) of the Cluster of Excellence PRISMA$^+$ (Project ID 390831469), for its hospitality and support.

\bibliography{refs.bib}{}

\begin{thebibliography}{93}%
\makeatletter
\providecommand \@ifxundefined [1]{%
 \@ifx{#1\undefined}
}%
\providecommand \@ifnum [1]{%
 \ifnum #1\expandafter \@firstoftwo
 \else \expandafter \@secondoftwo
 \fi
}%
\providecommand \@ifx [1]{%
 \ifx #1\expandafter \@firstoftwo
 \else \expandafter \@secondoftwo
 \fi
}%
\providecommand \natexlab [1]{#1}%
\providecommand \enquote  [1]{``#1''}%
\providecommand \bibnamefont  [1]{#1}%
\providecommand \bibfnamefont [1]{#1}%
\providecommand \citenamefont [1]{#1}%
\providecommand \href@noop [0]{\@secondoftwo}%
\providecommand \href [0]{\begingroup \@sanitize@url \@href}%
\providecommand \@href[1]{\@@startlink{#1}\@@href}%
\providecommand \@@href[1]{\endgroup#1\@@endlink}%
\providecommand \@sanitize@url [0]{\catcode `\\12\catcode `\$12\catcode
  `\&12\catcode `\#12\catcode `\^12\catcode `\_12\catcode `\%12\relax}%
\providecommand \@@startlink[1]{}%
\providecommand \@@endlink[0]{}%
\providecommand \url  [0]{\begingroup\@sanitize@url \@url }%
\providecommand \@url [1]{\endgroup\@href {#1}{\urlprefix }}%
\providecommand \urlprefix  [0]{URL }%
\providecommand \Eprint [0]{\href }%
\providecommand \doibase [0]{http://dx.doi.org/}%
\providecommand \selectlanguage [0]{\@gobble}%
\providecommand \bibinfo  [0]{\@secondoftwo}%
\providecommand \bibfield  [0]{\@secondoftwo}%
\providecommand \translation [1]{[#1]}%
\providecommand \BibitemOpen [0]{}%
\providecommand \bibitemStop [0]{}%
\providecommand \bibitemNoStop [0]{.\EOS\space}%
\providecommand \EOS [0]{\spacefactor3000\relax}%
\providecommand \BibitemShut  [1]{\csname bibitem#1\endcsname}%
\let\auto@bib@innerbib\@empty
\bibitem [{\citenamefont {Andres}\ and\ \citenamefont
  {Holguin}(2024)}]{Andres:2024pyz}%
  \BibitemOpen
  \bibfield  {author} {\bibinfo {author} {\bibfnamefont {C.}~\bibnamefont
  {Andres}}\ and\ \bibinfo {author} {\bibfnamefont {J.}~\bibnamefont
  {Holguin}},\ }\href@noop {} {\  (\bibinfo {year} {2024})},\ \Eprint
  {http://arxiv.org/abs/2409.07526} {arXiv:2409.07526 [hep-ph]} \BibitemShut
  {NoStop}%
\bibitem [{\citenamefont {Viinikainen}(2024)}]{talkEEC}%
  \BibitemOpen
  \bibfield  {author} {\bibinfo {author} {\bibfnamefont {J.}~\bibnamefont
  {Viinikainen}},\ }\href
  {https://indico.mitp.uni-mainz.de/event/358/contributions/4984/} {\enquote
  {\bibinfo {title} {Energy-energy correlators from {P}b{P}b and pp collisions
  at 5.02 {T}e{V} with {CMS}},}\ } (\bibinfo {year} {July 2024}),\ \bibinfo
  {note} {{E}nergy Correlators at the Collider Frontier, MITP, Mainz
  (Germany)}\BibitemShut {NoStop}%
\bibitem [{CMS(2024)}]{CMS-PAS-HIN-23-004}%
  \BibitemOpen
  \href {https://cds.cern.ch/record/2906425} {\emph {\bibinfo {title}
  {{Energy-energy correlators from PbPb and pp collisions at 5.02 TeV}}}},\
  \bibinfo {type} {Tech. Rep.}\ (\bibinfo  {institution} {CMS-PAS-HIN-23-004,
  CERN},\ \bibinfo {address} {Geneva},\ \bibinfo {year} {2024})\BibitemShut
  {NoStop}%
\bibitem [{\citenamefont {Busza}\ \emph {et~al.}(2018)\citenamefont {Busza},
  \citenamefont {Rajagopal},\ and\ \citenamefont {van~der
  Schee}}]{Busza:2018rrf}%
  \BibitemOpen
  \bibfield  {author} {\bibinfo {author} {\bibfnamefont {W.}~\bibnamefont
  {Busza}}, \bibinfo {author} {\bibfnamefont {K.}~\bibnamefont {Rajagopal}}, \
  and\ \bibinfo {author} {\bibfnamefont {W.}~\bibnamefont {van~der Schee}},\
  }\href {\doibase 10.1146/annurev-nucl-101917-020852} {\bibfield  {journal}
  {\bibinfo  {journal} {Ann. Rev. Nucl. Part. Sci.}\ }\textbf {\bibinfo
  {volume} {68}},\ \bibinfo {pages} {339} (\bibinfo {year} {2018})},\ \Eprint
  {http://arxiv.org/abs/1802.04801} {arXiv:1802.04801 [hep-ph]} \BibitemShut
  {NoStop}%
\bibitem [{\citenamefont {Cao}\ and\ \citenamefont {Wang}(2021)}]{Cao:2020wlm}%
  \BibitemOpen
  \bibfield  {author} {\bibinfo {author} {\bibfnamefont {S.}~\bibnamefont
  {Cao}}\ and\ \bibinfo {author} {\bibfnamefont {X.-N.}\ \bibnamefont {Wang}},\
  }\href {\doibase 10.1088/1361-6633/abc22b} {\bibfield  {journal} {\bibinfo
  {journal} {Rept. Prog. Phys.}\ }\textbf {\bibinfo {volume} {84}},\ \bibinfo
  {pages} {024301} (\bibinfo {year} {2021})},\ \Eprint
  {http://arxiv.org/abs/2002.04028} {arXiv:2002.04028 [hep-ph]} \BibitemShut
  {NoStop}%
\bibitem [{\citenamefont {Arslandok}\ \emph {et~al.}(2023)\citenamefont
  {Arslandok} \emph {et~al.}}]{Arslandok:2023utm}%
  \BibitemOpen
  \bibfield  {author} {\bibinfo {author} {\bibfnamefont {M.}~\bibnamefont
  {Arslandok}} \emph {et~al.},\ }\href@noop {} {\  (\bibinfo {year} {2023})},\
  \Eprint {http://arxiv.org/abs/2303.17254} {arXiv:2303.17254 [nucl-ex]}
  \BibitemShut {NoStop}%
\bibitem [{\citenamefont {Hayrapetyan}\ \emph
  {et~al.}(2024{\natexlab{a}})\citenamefont {Hayrapetyan} \emph
  {et~al.}}]{CMS:2024krd}%
  \BibitemOpen
  \bibfield  {author} {\bibinfo {author} {\bibfnamefont {A.}~\bibnamefont
  {Hayrapetyan}} \emph {et~al.} (\bibinfo {collaboration} {CMS}),\ }\href@noop
  {} {\  (\bibinfo {year} {2024}{\natexlab{a}})},\ \Eprint
  {http://arxiv.org/abs/2405.10785} {arXiv:2405.10785 [nucl-ex]} \BibitemShut
  {NoStop}%
\bibitem [{\citenamefont {Acharya}\ \emph {et~al.}(2024)\citenamefont {Acharya}
  \emph {et~al.}}]{ALICE:2022wpn}%
  \BibitemOpen
  \bibfield  {author} {\bibinfo {author} {\bibfnamefont {S.}~\bibnamefont
  {Acharya}} \emph {et~al.} (\bibinfo {collaboration} {ALICE}),\ }\href
  {\doibase 10.1140/epjc/s10052-024-12935-y} {\bibfield  {journal} {\bibinfo
  {journal} {Eur. Phys. J. C}\ }\textbf {\bibinfo {volume} {84}},\ \bibinfo
  {pages} {813} (\bibinfo {year} {2024})},\ \Eprint
  {http://arxiv.org/abs/2211.04384} {arXiv:2211.04384 [nucl-ex]} \BibitemShut
  {NoStop}%
\bibitem [{\citenamefont {Apolin\'ario}\ \emph {et~al.}(2022)\citenamefont
  {Apolin\'ario}, \citenamefont {Lee},\ and\ \citenamefont
  {Winn}}]{Apolinario:2022vzg}%
  \BibitemOpen
  \bibfield  {author} {\bibinfo {author} {\bibfnamefont {L.}~\bibnamefont
  {Apolin\'ario}}, \bibinfo {author} {\bibfnamefont {Y.-J.}\ \bibnamefont
  {Lee}}, \ and\ \bibinfo {author} {\bibfnamefont {M.}~\bibnamefont {Winn}},\
  }\href {\doibase 10.1016/j.ppnp.2022.103990} {\bibfield  {journal} {\bibinfo
  {journal} {Prog. Part. Nucl. Phys.}\ }\textbf {\bibinfo {volume} {127}},\
  \bibinfo {pages} {103990} (\bibinfo {year} {2022})},\ \Eprint
  {http://arxiv.org/abs/2203.16352} {arXiv:2203.16352 [hep-ph]} \BibitemShut
  {NoStop}%
\bibitem [{\citenamefont {Cunqueiro}\ and\ \citenamefont
  {Sickles}(2022)}]{Cunqueiro:2021wls}%
  \BibitemOpen
  \bibfield  {author} {\bibinfo {author} {\bibfnamefont {L.}~\bibnamefont
  {Cunqueiro}}\ and\ \bibinfo {author} {\bibfnamefont {A.~M.}\ \bibnamefont
  {Sickles}},\ }\href {\doibase 10.1016/j.ppnp.2022.103940} {\bibfield
  {journal} {\bibinfo  {journal} {Prog. Part. Nucl. Phys.}\ }\textbf {\bibinfo
  {volume} {124}},\ \bibinfo {pages} {103940} (\bibinfo {year} {2022})},\
  \Eprint {http://arxiv.org/abs/2110.14490} {arXiv:2110.14490 [nucl-ex]}
  \BibitemShut {NoStop}%
\bibitem [{\citenamefont {Connors}\ \emph {et~al.}(2018)\citenamefont
  {Connors}, \citenamefont {Nattrass}, \citenamefont {Reed},\ and\
  \citenamefont {Salur}}]{Connors:2017ptx}%
  \BibitemOpen
  \bibfield  {author} {\bibinfo {author} {\bibfnamefont {M.}~\bibnamefont
  {Connors}}, \bibinfo {author} {\bibfnamefont {C.}~\bibnamefont {Nattrass}},
  \bibinfo {author} {\bibfnamefont {R.}~\bibnamefont {Reed}}, \ and\ \bibinfo
  {author} {\bibfnamefont {S.}~\bibnamefont {Salur}},\ }\href {\doibase
  10.1103/RevModPhys.90.025005} {\bibfield  {journal} {\bibinfo  {journal}
  {Rev. Mod. Phys.}\ }\textbf {\bibinfo {volume} {90}},\ \bibinfo {pages}
  {025005} (\bibinfo {year} {2018})},\ \Eprint
  {http://arxiv.org/abs/1705.01974} {arXiv:1705.01974 [nucl-ex]} \BibitemShut
  {NoStop}%
\bibitem [{\citenamefont {Basham}\ \emph
  {et~al.}(1978{\natexlab{a}})\citenamefont {Basham}, \citenamefont {Brown},
  \citenamefont {Ellis},\ and\ \citenamefont {Love}}]{Basham:1977iq}%
  \BibitemOpen
  \bibfield  {author} {\bibinfo {author} {\bibfnamefont {C.~L.}\ \bibnamefont
  {Basham}}, \bibinfo {author} {\bibfnamefont {L.~S.}\ \bibnamefont {Brown}},
  \bibinfo {author} {\bibfnamefont {S.~D.}\ \bibnamefont {Ellis}}, \ and\
  \bibinfo {author} {\bibfnamefont {S.~T.}\ \bibnamefont {Love}},\ }\href
  {\doibase 10.1103/PhysRevD.17.2298} {\bibfield  {journal} {\bibinfo
  {journal} {Phys. Rev. D}\ }\textbf {\bibinfo {volume} {17}},\ \bibinfo
  {pages} {2298} (\bibinfo {year} {1978}{\natexlab{a}})}\BibitemShut {NoStop}%
\bibitem [{\citenamefont {Basham}\ \emph
  {et~al.}(1978{\natexlab{b}})\citenamefont {Basham}, \citenamefont {Brown},
  \citenamefont {Ellis},\ and\ \citenamefont {Love}}]{Basham:1978bw}%
  \BibitemOpen
  \bibfield  {author} {\bibinfo {author} {\bibfnamefont {C.}~\bibnamefont
  {Basham}}, \bibinfo {author} {\bibfnamefont {L.~S.}\ \bibnamefont {Brown}},
  \bibinfo {author} {\bibfnamefont {S.~D.}\ \bibnamefont {Ellis}}, \ and\
  \bibinfo {author} {\bibfnamefont {S.~T.}\ \bibnamefont {Love}},\ }\href
  {\doibase 10.1103/PhysRevLett.41.1585} {\bibfield  {journal} {\bibinfo
  {journal} {Phys. Rev. Lett.}\ }\textbf {\bibinfo {volume} {41}},\ \bibinfo
  {pages} {1585} (\bibinfo {year} {1978}{\natexlab{b}})}\BibitemShut {NoStop}%
\bibitem [{\citenamefont {Basham}\ \emph
  {et~al.}(1979{\natexlab{a}})\citenamefont {Basham}, \citenamefont {Brown},
  \citenamefont {Ellis},\ and\ \citenamefont {Love}}]{Basham:1978zq}%
  \BibitemOpen
  \bibfield  {author} {\bibinfo {author} {\bibfnamefont {C.}~\bibnamefont
  {Basham}}, \bibinfo {author} {\bibfnamefont {L.}~\bibnamefont {Brown}},
  \bibinfo {author} {\bibfnamefont {S.}~\bibnamefont {Ellis}}, \ and\ \bibinfo
  {author} {\bibfnamefont {S.}~\bibnamefont {Love}},\ }\href {\doibase
  10.1103/PhysRevD.19.2018} {\bibfield  {journal} {\bibinfo  {journal} {Phys.
  Rev. D}\ }\textbf {\bibinfo {volume} {19}},\ \bibinfo {pages} {2018}
  (\bibinfo {year} {1979}{\natexlab{a}})}\BibitemShut {NoStop}%
\bibitem [{\citenamefont {Basham}\ \emph
  {et~al.}(1979{\natexlab{b}})\citenamefont {Basham}, \citenamefont {Brown},
  \citenamefont {Ellis},\ and\ \citenamefont {Love}}]{Basham:1979gh}%
  \BibitemOpen
  \bibfield  {author} {\bibinfo {author} {\bibfnamefont {C.~L.}\ \bibnamefont
  {Basham}}, \bibinfo {author} {\bibfnamefont {L.~S.}\ \bibnamefont {Brown}},
  \bibinfo {author} {\bibfnamefont {S.~D.}\ \bibnamefont {Ellis}}, \ and\
  \bibinfo {author} {\bibfnamefont {S.~T.}\ \bibnamefont {Love}},\ }\href
  {\doibase 10.1016/0370-2693(79)90601-4} {\bibfield  {journal} {\bibinfo
  {journal} {Phys. Lett. B}\ }\textbf {\bibinfo {volume} {85}},\ \bibinfo
  {pages} {297} (\bibinfo {year} {1979}{\natexlab{b}})}\BibitemShut {NoStop}%
\bibitem [{\citenamefont {Hofman}\ and\ \citenamefont
  {Maldacena}(2008)}]{Hofman:2008ar}%
  \BibitemOpen
  \bibfield  {author} {\bibinfo {author} {\bibfnamefont {D.~M.}\ \bibnamefont
  {Hofman}}\ and\ \bibinfo {author} {\bibfnamefont {J.}~\bibnamefont
  {Maldacena}},\ }\href {\doibase 10.1088/1126-6708/2008/05/012} {\bibfield
  {journal} {\bibinfo  {journal} {JHEP}\ }\textbf {\bibinfo {volume} {05}},\
  \bibinfo {pages} {012} (\bibinfo {year} {2008})},\ \Eprint
  {http://arxiv.org/abs/0803.1467} {arXiv:0803.1467 [hep-th]} \BibitemShut
  {NoStop}%
\bibitem [{\citenamefont {Chang}\ \emph
  {et~al.}(2013{\natexlab{a}})\citenamefont {Chang}, \citenamefont {Procura},
  \citenamefont {Thaler},\ and\ \citenamefont {Waalewijn}}]{Chang:2013rca}%
  \BibitemOpen
  \bibfield  {author} {\bibinfo {author} {\bibfnamefont {H.-M.}\ \bibnamefont
  {Chang}}, \bibinfo {author} {\bibfnamefont {M.}~\bibnamefont {Procura}},
  \bibinfo {author} {\bibfnamefont {J.}~\bibnamefont {Thaler}}, \ and\ \bibinfo
  {author} {\bibfnamefont {W.~J.}\ \bibnamefont {Waalewijn}},\ }\href {\doibase
  10.1103/PhysRevLett.111.102002} {\bibfield  {journal} {\bibinfo  {journal}
  {Phys. Rev. Lett.}\ }\textbf {\bibinfo {volume} {111}},\ \bibinfo {pages}
  {102002} (\bibinfo {year} {2013}{\natexlab{a}})},\ \Eprint
  {http://arxiv.org/abs/1303.6637} {arXiv:1303.6637 [hep-ph]} \BibitemShut
  {NoStop}%
\bibitem [{\citenamefont {Chang}\ \emph
  {et~al.}(2013{\natexlab{b}})\citenamefont {Chang}, \citenamefont {Procura},
  \citenamefont {Thaler},\ and\ \citenamefont {Waalewijn}}]{Chang:2013iba}%
  \BibitemOpen
  \bibfield  {author} {\bibinfo {author} {\bibfnamefont {H.-M.}\ \bibnamefont
  {Chang}}, \bibinfo {author} {\bibfnamefont {M.}~\bibnamefont {Procura}},
  \bibinfo {author} {\bibfnamefont {J.}~\bibnamefont {Thaler}}, \ and\ \bibinfo
  {author} {\bibfnamefont {W.~J.}\ \bibnamefont {Waalewijn}},\ }\href {\doibase
  10.1103/PhysRevD.88.034030} {\bibfield  {journal} {\bibinfo  {journal} {Phys.
  Rev.}\ }\textbf {\bibinfo {volume} {D88}},\ \bibinfo {pages} {034030}
  (\bibinfo {year} {2013}{\natexlab{b}})},\ \Eprint
  {http://arxiv.org/abs/1306.6630} {arXiv:1306.6630 [hep-ph]} \BibitemShut
  {NoStop}%
\bibitem [{\citenamefont {Li}\ \emph {et~al.}(2022)\citenamefont {Li},
  \citenamefont {Moult}, \citenamefont {van Velzen}, \citenamefont
  {Waalewijn},\ and\ \citenamefont {Zhu}}]{Li:2021zcf}%
  \BibitemOpen
  \bibfield  {author} {\bibinfo {author} {\bibfnamefont {Y.}~\bibnamefont
  {Li}}, \bibinfo {author} {\bibfnamefont {I.}~\bibnamefont {Moult}}, \bibinfo
  {author} {\bibfnamefont {S.~S.}\ \bibnamefont {van Velzen}}, \bibinfo
  {author} {\bibfnamefont {W.~J.}\ \bibnamefont {Waalewijn}}, \ and\ \bibinfo
  {author} {\bibfnamefont {H.~X.}\ \bibnamefont {Zhu}},\ }\href {\doibase
  10.1103/PhysRevLett.128.182001} {\bibfield  {journal} {\bibinfo  {journal}
  {Phys. Rev. Lett.}\ }\textbf {\bibinfo {volume} {128}},\ \bibinfo {pages}
  {182001} (\bibinfo {year} {2022})},\ \Eprint
  {http://arxiv.org/abs/2108.01674} {arXiv:2108.01674 [hep-ph]} \BibitemShut
  {NoStop}%
\bibitem [{\citenamefont {Chen}\ \emph {et~al.}(2022)\citenamefont {Chen},
  \citenamefont {Jaarsma}, \citenamefont {Li}, \citenamefont {Moult},
  \citenamefont {Waalewijn},\ and\ \citenamefont {Zhu}}]{Chen:2022muj}%
  \BibitemOpen
  \bibfield  {author} {\bibinfo {author} {\bibfnamefont {H.}~\bibnamefont
  {Chen}}, \bibinfo {author} {\bibfnamefont {M.}~\bibnamefont {Jaarsma}},
  \bibinfo {author} {\bibfnamefont {Y.}~\bibnamefont {Li}}, \bibinfo {author}
  {\bibfnamefont {I.}~\bibnamefont {Moult}}, \bibinfo {author} {\bibfnamefont
  {W.~J.}\ \bibnamefont {Waalewijn}}, \ and\ \bibinfo {author} {\bibfnamefont
  {H.~X.}\ \bibnamefont {Zhu}},\ }\href@noop {} {\  (\bibinfo {year} {2022})},\
  \Eprint {http://arxiv.org/abs/2210.10061} {arXiv:2210.10061 [hep-ph]}
  \BibitemShut {NoStop}%
\bibitem [{\citenamefont {Chen}\ \emph {et~al.}(2023)\citenamefont {Chen},
  \citenamefont {Jaarsma}, \citenamefont {Li}, \citenamefont {Moult},
  \citenamefont {Waalewijn},\ and\ \citenamefont {Zhu}}]{Chen:2022pdu}%
  \BibitemOpen
  \bibfield  {author} {\bibinfo {author} {\bibfnamefont {H.}~\bibnamefont
  {Chen}}, \bibinfo {author} {\bibfnamefont {M.}~\bibnamefont {Jaarsma}},
  \bibinfo {author} {\bibfnamefont {Y.}~\bibnamefont {Li}}, \bibinfo {author}
  {\bibfnamefont {I.}~\bibnamefont {Moult}}, \bibinfo {author} {\bibfnamefont
  {W.~J.}\ \bibnamefont {Waalewijn}}, \ and\ \bibinfo {author} {\bibfnamefont
  {H.~X.}\ \bibnamefont {Zhu}},\ }\href {\doibase 10.1007/JHEP07(2023)185}
  {\bibfield  {journal} {\bibinfo  {journal} {JHEP}\ }\textbf {\bibinfo
  {volume} {07}},\ \bibinfo {pages} {185} (\bibinfo {year} {2023})},\ \Eprint
  {http://arxiv.org/abs/2210.10058} {arXiv:2210.10058 [hep-ph]} \BibitemShut
  {NoStop}%
\bibitem [{\citenamefont {Jaarsma}\ \emph {et~al.}(2022)\citenamefont
  {Jaarsma}, \citenamefont {Li}, \citenamefont {Moult}, \citenamefont
  {Waalewijn},\ and\ \citenamefont {Zhu}}]{Jaarsma:2022kdd}%
  \BibitemOpen
  \bibfield  {author} {\bibinfo {author} {\bibfnamefont {M.}~\bibnamefont
  {Jaarsma}}, \bibinfo {author} {\bibfnamefont {Y.}~\bibnamefont {Li}},
  \bibinfo {author} {\bibfnamefont {I.}~\bibnamefont {Moult}}, \bibinfo
  {author} {\bibfnamefont {W.}~\bibnamefont {Waalewijn}}, \ and\ \bibinfo
  {author} {\bibfnamefont {H.~X.}\ \bibnamefont {Zhu}},\ }\href {\doibase
  10.1007/JHEP06(2022)139} {\bibfield  {journal} {\bibinfo  {journal} {JHEP}\
  }\textbf {\bibinfo {volume} {06}},\ \bibinfo {pages} {139} (\bibinfo {year}
  {2022})},\ \Eprint {http://arxiv.org/abs/2201.05166} {arXiv:2201.05166
  [hep-ph]} \BibitemShut {NoStop}%
\bibitem [{\citenamefont {Jaarsma}\ \emph {et~al.}(2023)\citenamefont
  {Jaarsma}, \citenamefont {Li}, \citenamefont {Moult}, \citenamefont
  {Waalewijn},\ and\ \citenamefont {Zhu}}]{Jaarsma:2023ell}%
  \BibitemOpen
  \bibfield  {author} {\bibinfo {author} {\bibfnamefont {M.}~\bibnamefont
  {Jaarsma}}, \bibinfo {author} {\bibfnamefont {Y.}~\bibnamefont {Li}},
  \bibinfo {author} {\bibfnamefont {I.}~\bibnamefont {Moult}}, \bibinfo
  {author} {\bibfnamefont {W.~J.}\ \bibnamefont {Waalewijn}}, \ and\ \bibinfo
  {author} {\bibfnamefont {H.~X.}\ \bibnamefont {Zhu}},\ }\href {\doibase
  10.1007/JHEP12(2023)087} {\bibfield  {journal} {\bibinfo  {journal} {JHEP}\
  }\textbf {\bibinfo {volume} {12}},\ \bibinfo {pages} {087} (\bibinfo {year}
  {2023})},\ \Eprint {http://arxiv.org/abs/2307.15739} {arXiv:2307.15739
  [hep-ph]} \BibitemShut {NoStop}%
\bibitem [{\citenamefont {Barata}\ and\ \citenamefont
  {Szafron}(2024)}]{Barata:2024nqo}%
  \BibitemOpen
  \bibfield  {author} {\bibinfo {author} {\bibfnamefont {J.~a.}\ \bibnamefont
  {Barata}}\ and\ \bibinfo {author} {\bibfnamefont {R.}~\bibnamefont
  {Szafron}},\ }\href {\doibase 10.1103/PhysRevD.110.L031501} {\bibfield
  {journal} {\bibinfo  {journal} {Phys. Rev. D}\ }\textbf {\bibinfo {volume}
  {110}},\ \bibinfo {pages} {L031501} (\bibinfo {year} {2024})},\ \Eprint
  {http://arxiv.org/abs/2401.04164} {arXiv:2401.04164 [hep-ph]} \BibitemShut
  {NoStop}%
\bibitem [{\citenamefont {Andres}\ \emph {et~al.}(2024)\citenamefont {Andres},
  \citenamefont {Dominguez}, \citenamefont {Holguin}, \citenamefont {Marquet},\
  and\ \citenamefont {Moult}}]{Andres:2024ksi}%
  \BibitemOpen
  \bibfield  {author} {\bibinfo {author} {\bibfnamefont {C.}~\bibnamefont
  {Andres}}, \bibinfo {author} {\bibfnamefont {F.}~\bibnamefont {Dominguez}},
  \bibinfo {author} {\bibfnamefont {J.}~\bibnamefont {Holguin}}, \bibinfo
  {author} {\bibfnamefont {C.}~\bibnamefont {Marquet}}, \ and\ \bibinfo
  {author} {\bibfnamefont {I.}~\bibnamefont {Moult}},\ }\href@noop {} {\
  (\bibinfo {year} {2024})},\ \Eprint {http://arxiv.org/abs/2407.07936}
  {arXiv:2407.07936 [hep-ph]} \BibitemShut {NoStop}%
\bibitem [{\citenamefont {Andres}\ \emph
  {et~al.}(2023{\natexlab{a}})\citenamefont {Andres}, \citenamefont
  {Dominguez}, \citenamefont {Kunnawalkam~Elayavalli}, \citenamefont {Holguin},
  \citenamefont {Marquet},\ and\ \citenamefont {Moult}}]{Andres:2022ovj}%
  \BibitemOpen
  \bibfield  {author} {\bibinfo {author} {\bibfnamefont {C.}~\bibnamefont
  {Andres}}, \bibinfo {author} {\bibfnamefont {F.}~\bibnamefont {Dominguez}},
  \bibinfo {author} {\bibfnamefont {R.}~\bibnamefont {Kunnawalkam~Elayavalli}},
  \bibinfo {author} {\bibfnamefont {J.}~\bibnamefont {Holguin}}, \bibinfo
  {author} {\bibfnamefont {C.}~\bibnamefont {Marquet}}, \ and\ \bibinfo
  {author} {\bibfnamefont {I.}~\bibnamefont {Moult}},\ }\href {\doibase
  10.1103/PhysRevLett.130.262301} {\bibfield  {journal} {\bibinfo  {journal}
  {Phys. Rev. Lett.}\ }\textbf {\bibinfo {volume} {130}},\ \bibinfo {pages}
  {262301} (\bibinfo {year} {2023}{\natexlab{a}})},\ \Eprint
  {http://arxiv.org/abs/2209.11236} {arXiv:2209.11236 [hep-ph]} \BibitemShut
  {NoStop}%
\bibitem [{\citenamefont {Andres}\ \emph
  {et~al.}(2023{\natexlab{b}})\citenamefont {Andres}, \citenamefont
  {Dominguez}, \citenamefont {Holguin}, \citenamefont {Marquet},\ and\
  \citenamefont {Moult}}]{Andres:2023xwr}%
  \BibitemOpen
  \bibfield  {author} {\bibinfo {author} {\bibfnamefont {C.}~\bibnamefont
  {Andres}}, \bibinfo {author} {\bibfnamefont {F.}~\bibnamefont {Dominguez}},
  \bibinfo {author} {\bibfnamefont {J.}~\bibnamefont {Holguin}}, \bibinfo
  {author} {\bibfnamefont {C.}~\bibnamefont {Marquet}}, \ and\ \bibinfo
  {author} {\bibfnamefont {I.}~\bibnamefont {Moult}},\ }\href {\doibase
  10.1007/JHEP09(2023)088} {\bibfield  {journal} {\bibinfo  {journal} {JHEP}\
  }\textbf {\bibinfo {volume} {09}},\ \bibinfo {pages} {088} (\bibinfo {year}
  {2023}{\natexlab{b}})},\ \Eprint {http://arxiv.org/abs/2303.03413}
  {arXiv:2303.03413 [hep-ph]} \BibitemShut {NoStop}%
\bibitem [{\citenamefont {Yang}\ \emph {et~al.}(2024)\citenamefont {Yang},
  \citenamefont {He}, \citenamefont {Moult},\ and\ \citenamefont
  {Wang}}]{Yang:2023dwc}%
  \BibitemOpen
  \bibfield  {author} {\bibinfo {author} {\bibfnamefont {Z.}~\bibnamefont
  {Yang}}, \bibinfo {author} {\bibfnamefont {Y.}~\bibnamefont {He}}, \bibinfo
  {author} {\bibfnamefont {I.}~\bibnamefont {Moult}}, \ and\ \bibinfo {author}
  {\bibfnamefont {X.-N.}\ \bibnamefont {Wang}},\ }\href {\doibase
  10.1103/PhysRevLett.132.011901} {\bibfield  {journal} {\bibinfo  {journal}
  {Phys. Rev. Lett.}\ }\textbf {\bibinfo {volume} {132}},\ \bibinfo {pages}
  {011901} (\bibinfo {year} {2024})},\ \Eprint
  {http://arxiv.org/abs/2310.01500} {arXiv:2310.01500 [hep-ph]} \BibitemShut
  {NoStop}%
\bibitem [{\citenamefont {Bossi}\ \emph {et~al.}(2024)\citenamefont {Bossi},
  \citenamefont {Kudinoor}, \citenamefont {Moult}, \citenamefont {Pablos},
  \citenamefont {Rai},\ and\ \citenamefont {Rajagopal}}]{Bossi:2024qho}%
  \BibitemOpen
  \bibfield  {author} {\bibinfo {author} {\bibfnamefont {H.}~\bibnamefont
  {Bossi}}, \bibinfo {author} {\bibfnamefont {A.~S.}\ \bibnamefont {Kudinoor}},
  \bibinfo {author} {\bibfnamefont {I.}~\bibnamefont {Moult}}, \bibinfo
  {author} {\bibfnamefont {D.}~\bibnamefont {Pablos}}, \bibinfo {author}
  {\bibfnamefont {A.}~\bibnamefont {Rai}}, \ and\ \bibinfo {author}
  {\bibfnamefont {K.}~\bibnamefont {Rajagopal}},\ }\href@noop {} {\  (\bibinfo
  {year} {2024})},\ \Eprint {http://arxiv.org/abs/2407.13818} {arXiv:2407.13818
  [hep-ph]} \BibitemShut {NoStop}%
\bibitem [{\citenamefont {Barata}\ \emph {et~al.}(2023)\citenamefont {Barata},
  \citenamefont {Caucal}, \citenamefont {Soto-Ontoso},\ and\ \citenamefont
  {Szafron}}]{Barata:2023bhh}%
  \BibitemOpen
  \bibfield  {author} {\bibinfo {author} {\bibfnamefont {J.~a.}\ \bibnamefont
  {Barata}}, \bibinfo {author} {\bibfnamefont {P.}~\bibnamefont {Caucal}},
  \bibinfo {author} {\bibfnamefont {A.}~\bibnamefont {Soto-Ontoso}}, \ and\
  \bibinfo {author} {\bibfnamefont {R.}~\bibnamefont {Szafron}},\ }\href@noop
  {} {\  (\bibinfo {year} {2023})},\ \Eprint {http://arxiv.org/abs/2312.12527}
  {arXiv:2312.12527 [hep-ph]} \BibitemShut {NoStop}%
\bibitem [{\citenamefont {Hayrapetyan}\ \emph
  {et~al.}(2024{\natexlab{b}})\citenamefont {Hayrapetyan} \emph
  {et~al.}}]{CMS:2024mlf}%
  \BibitemOpen
  \bibfield  {author} {\bibinfo {author} {\bibfnamefont {A.}~\bibnamefont
  {Hayrapetyan}} \emph {et~al.} (\bibinfo {collaboration} {CMS}),\ }\href
  {\doibase 10.1103/PhysRevLett.133.071903} {\bibfield  {journal} {\bibinfo
  {journal} {Phys. Rev. Lett.}\ }\textbf {\bibinfo {volume} {133}},\ \bibinfo
  {pages} {071903} (\bibinfo {year} {2024}{\natexlab{b}})},\ \Eprint
  {http://arxiv.org/abs/2402.13864} {arXiv:2402.13864 [hep-ex]} \BibitemShut
  {NoStop}%
\bibitem [{\citenamefont {Cruz-Torres}()}]{talk2}%
  \BibitemOpen
  \bibfield  {author} {\bibinfo {author} {\bibfnamefont {R.}~\bibnamefont
  {Cruz-Torres}},\ }\href@noop {} {\bibinfo  {journal} {Measurement of the
  angle between jet axes and energy-energy correlators with ALICE, Hard Probes
  2023}\ }\BibitemShut {NoStop}%
\bibitem [{\citenamefont {Lu}()}]{talk4}%
  \BibitemOpen
\bibfield  {journal} {  }\bibfield  {author} {\bibinfo {author} {\bibfnamefont
  {C.}~\bibnamefont {Lu}},\ }\href@noop {} {\ }\bibinfo {note} {15th
  International Workshop on Boosted Object Phenomenology, Reconstruction,
  Measurements, and Searches at Colliders (Boost 2023)}\BibitemShut {NoStop}%
\bibitem [{\citenamefont {Fan}({\natexlab{a}})}]{talk6}%
  \BibitemOpen
  \bibfield  {author} {\bibinfo {author} {\bibfnamefont {W.}~\bibnamefont
  {Fan}},\ }\href@noop {} {\  ({\natexlab{a}})},\ \bibinfo {note} {xXXth
  International Conference on Ultra-relativistic Nucleus-Nucleus Collisions
  (Quark Matter 2023)}\BibitemShut {NoStop}%
\bibitem [{\citenamefont {Tamis}(2024)}]{Tamis:2023guc}%
  \BibitemOpen
  \bibfield  {author} {\bibinfo {author} {\bibfnamefont {A.}~\bibnamefont
  {Tamis}} (\bibinfo {collaboration} {STAR}),\ }\href {\doibase
  10.22323/1.438.0175} {\bibfield  {journal} {\bibinfo  {journal} {PoS}\
  }\textbf {\bibinfo {volume} {HardProbes2023}},\ \bibinfo {pages} {175}
  (\bibinfo {year} {2024})},\ \Eprint {http://arxiv.org/abs/2309.05761}
  {arXiv:2309.05761 [hep-ex]} \BibitemShut {NoStop}%
\bibitem [{\citenamefont {Baier}\ \emph {et~al.}(2001)\citenamefont {Baier},
  \citenamefont {Dokshitzer}, \citenamefont {Mueller},\ and\ \citenamefont
  {Schiff}}]{Baier:2001yt}%
  \BibitemOpen
  \bibfield  {author} {\bibinfo {author} {\bibfnamefont {R.}~\bibnamefont
  {Baier}}, \bibinfo {author} {\bibfnamefont {Y.~L.}\ \bibnamefont
  {Dokshitzer}}, \bibinfo {author} {\bibfnamefont {A.~H.}\ \bibnamefont
  {Mueller}}, \ and\ \bibinfo {author} {\bibfnamefont {D.}~\bibnamefont
  {Schiff}},\ }\href {\doibase 10.1088/1126-6708/2001/09/033} {\bibfield
  {journal} {\bibinfo  {journal} {JHEP}\ }\textbf {\bibinfo {volume} {09}},\
  \bibinfo {pages} {033} (\bibinfo {year} {2001})},\ \Eprint
  {http://arxiv.org/abs/hep-ph/0106347} {arXiv:hep-ph/0106347} \BibitemShut
  {NoStop}%
\bibitem [{\citenamefont {Zhang}\ \emph {et~al.}(2009)\citenamefont {Zhang},
  \citenamefont {Owens}, \citenamefont {Wang},\ and\ \citenamefont
  {Wang}}]{Zhang:2009rn}%
  \BibitemOpen
  \bibfield  {author} {\bibinfo {author} {\bibfnamefont {H.}~\bibnamefont
  {Zhang}}, \bibinfo {author} {\bibfnamefont {J.~F.}\ \bibnamefont {Owens}},
  \bibinfo {author} {\bibfnamefont {E.}~\bibnamefont {Wang}}, \ and\ \bibinfo
  {author} {\bibfnamefont {X.-N.}\ \bibnamefont {Wang}},\ }\href {\doibase
  10.1103/PhysRevLett.103.032302} {\bibfield  {journal} {\bibinfo  {journal}
  {Phys. Rev. Lett.}\ }\textbf {\bibinfo {volume} {103}},\ \bibinfo {pages}
  {032302} (\bibinfo {year} {2009})},\ \Eprint {http://arxiv.org/abs/0902.4000}
  {arXiv:0902.4000 [nucl-th]} \BibitemShut {NoStop}%
\bibitem [{\citenamefont {d'Enterria}(2010)}]{dEnterria:2009xfs}%
  \BibitemOpen
  \bibfield  {author} {\bibinfo {author} {\bibfnamefont {D.}~\bibnamefont
  {d'Enterria}},\ }\href {\doibase 10.1007/978-3-642-01539-7_16} {\bibfield
  {journal} {\bibinfo  {journal} {Landolt-Bornstein}\ }\textbf {\bibinfo
  {volume} {23}},\ \bibinfo {pages} {471} (\bibinfo {year} {2010})},\ \Eprint
  {http://arxiv.org/abs/0902.2011} {arXiv:0902.2011 [nucl-ex]} \BibitemShut
  {NoStop}%
\bibitem [{\citenamefont {Renk}(2013)}]{Renk:2012ve}%
  \BibitemOpen
  \bibfield  {author} {\bibinfo {author} {\bibfnamefont {T.}~\bibnamefont
  {Renk}},\ }\href {\doibase 10.1103/PhysRevC.88.054902} {\bibfield  {journal}
  {\bibinfo  {journal} {Phys. Rev. C}\ }\textbf {\bibinfo {volume} {88}},\
  \bibinfo {pages} {054902} (\bibinfo {year} {2013})},\ \Eprint
  {http://arxiv.org/abs/1212.0646} {arXiv:1212.0646 [hep-ph]} \BibitemShut
  {NoStop}%
\bibitem [{\citenamefont {Milhano}\ and\ \citenamefont
  {Zapp}(2016)}]{Milhano:2015mng}%
  \BibitemOpen
  \bibfield  {author} {\bibinfo {author} {\bibfnamefont {J.~G.}\ \bibnamefont
  {Milhano}}\ and\ \bibinfo {author} {\bibfnamefont {K.~C.}\ \bibnamefont
  {Zapp}},\ }\href {\doibase 10.1140/epjc/s10052-016-4130-9} {\bibfield
  {journal} {\bibinfo  {journal} {Eur. Phys. J. C}\ }\textbf {\bibinfo {volume}
  {76}},\ \bibinfo {pages} {288} (\bibinfo {year} {2016})},\ \Eprint
  {http://arxiv.org/abs/1512.08107} {arXiv:1512.08107 [hep-ph]} \BibitemShut
  {NoStop}%
\bibitem [{\citenamefont {Spousta}\ and\ \citenamefont
  {Cole}(2016)}]{Spousta:2015fca}%
  \BibitemOpen
  \bibfield  {author} {\bibinfo {author} {\bibfnamefont {M.}~\bibnamefont
  {Spousta}}\ and\ \bibinfo {author} {\bibfnamefont {B.}~\bibnamefont {Cole}},\
  }\href {\doibase 10.1140/epjc/s10052-016-3896-0} {\bibfield  {journal}
  {\bibinfo  {journal} {Eur. Phys. J. C}\ }\textbf {\bibinfo {volume} {76}},\
  \bibinfo {pages} {50} (\bibinfo {year} {2016})},\ \Eprint
  {http://arxiv.org/abs/1504.05169} {arXiv:1504.05169 [hep-ph]} \BibitemShut
  {NoStop}%
\bibitem [{\citenamefont {Rajagopal}\ \emph {et~al.}(2016)\citenamefont
  {Rajagopal}, \citenamefont {Sadofyev},\ and\ \citenamefont {van~der
  Schee}}]{Rajagopal:2016uip}%
  \BibitemOpen
  \bibfield  {author} {\bibinfo {author} {\bibfnamefont {K.}~\bibnamefont
  {Rajagopal}}, \bibinfo {author} {\bibfnamefont {A.~V.}\ \bibnamefont
  {Sadofyev}}, \ and\ \bibinfo {author} {\bibfnamefont {W.}~\bibnamefont
  {van~der Schee}},\ }\href {\doibase 10.1103/PhysRevLett.116.211603}
  {\bibfield  {journal} {\bibinfo  {journal} {Phys. Rev. Lett.}\ }\textbf
  {\bibinfo {volume} {116}},\ \bibinfo {pages} {211603} (\bibinfo {year}
  {2016})},\ \Eprint {http://arxiv.org/abs/1602.04187} {arXiv:1602.04187
  [nucl-th]} \BibitemShut {NoStop}%
\bibitem [{\citenamefont {Casalderrey-Solana}\ \emph
  {et~al.}(2016)\citenamefont {Casalderrey-Solana}, \citenamefont {Gulhan},
  \citenamefont {Milhano}, \citenamefont {Pablos},\ and\ \citenamefont
  {Rajagopal}}]{Casalderrey-Solana:2015vaa}%
  \BibitemOpen
  \bibfield  {author} {\bibinfo {author} {\bibfnamefont {J.}~\bibnamefont
  {Casalderrey-Solana}}, \bibinfo {author} {\bibfnamefont {D.~C.}\ \bibnamefont
  {Gulhan}}, \bibinfo {author} {\bibfnamefont {J.~G.}\ \bibnamefont {Milhano}},
  \bibinfo {author} {\bibfnamefont {D.}~\bibnamefont {Pablos}}, \ and\ \bibinfo
  {author} {\bibfnamefont {K.}~\bibnamefont {Rajagopal}},\ }\href {\doibase
  10.1007/JHEP03(2016)053} {\bibfield  {journal} {\bibinfo  {journal} {JHEP}\
  }\textbf {\bibinfo {volume} {03}},\ \bibinfo {pages} {053} (\bibinfo {year}
  {2016})},\ \Eprint {http://arxiv.org/abs/1508.00815} {arXiv:1508.00815
  [hep-ph]} \BibitemShut {NoStop}%
\bibitem [{\citenamefont {Casalderrey-Solana}\ \emph
  {et~al.}(2017)\citenamefont {Casalderrey-Solana}, \citenamefont {Gulhan},
  \citenamefont {Milhano}, \citenamefont {Pablos},\ and\ \citenamefont
  {Rajagopal}}]{Casalderrey-Solana:2016jvj}%
  \BibitemOpen
  \bibfield  {author} {\bibinfo {author} {\bibfnamefont {J.}~\bibnamefont
  {Casalderrey-Solana}}, \bibinfo {author} {\bibfnamefont {D.}~\bibnamefont
  {Gulhan}}, \bibinfo {author} {\bibfnamefont {G.}~\bibnamefont {Milhano}},
  \bibinfo {author} {\bibfnamefont {D.}~\bibnamefont {Pablos}}, \ and\ \bibinfo
  {author} {\bibfnamefont {K.}~\bibnamefont {Rajagopal}},\ }\href {\doibase
  10.1007/JHEP03(2017)135} {\bibfield  {journal} {\bibinfo  {journal} {JHEP}\
  }\textbf {\bibinfo {volume} {03}},\ \bibinfo {pages} {135} (\bibinfo {year}
  {2017})},\ \Eprint {http://arxiv.org/abs/1609.05842} {arXiv:1609.05842
  [hep-ph]} \BibitemShut {NoStop}%
\bibitem [{\citenamefont {Caucal}\ \emph {et~al.}(2018)\citenamefont {Caucal},
  \citenamefont {Iancu}, \citenamefont {Mueller},\ and\ \citenamefont
  {Soyez}}]{Caucal:2018dla}%
  \BibitemOpen
  \bibfield  {author} {\bibinfo {author} {\bibfnamefont {P.}~\bibnamefont
  {Caucal}}, \bibinfo {author} {\bibfnamefont {E.}~\bibnamefont {Iancu}},
  \bibinfo {author} {\bibfnamefont {A.~H.}\ \bibnamefont {Mueller}}, \ and\
  \bibinfo {author} {\bibfnamefont {G.}~\bibnamefont {Soyez}},\ }\href
  {\doibase 10.1103/PhysRevLett.120.232001} {\bibfield  {journal} {\bibinfo
  {journal} {Phys. Rev. Lett.}\ }\textbf {\bibinfo {volume} {120}},\ \bibinfo
  {pages} {232001} (\bibinfo {year} {2018})},\ \Eprint
  {http://arxiv.org/abs/1801.09703} {arXiv:1801.09703 [hep-ph]} \BibitemShut
  {NoStop}%
\bibitem [{\citenamefont {Casalderrey-Solana}\ \emph
  {et~al.}(2019)\citenamefont {Casalderrey-Solana}, \citenamefont {Hulcher},
  \citenamefont {Milhano}, \citenamefont {Pablos},\ and\ \citenamefont
  {Rajagopal}}]{Casalderrey-Solana:2018wrw}%
  \BibitemOpen
  \bibfield  {author} {\bibinfo {author} {\bibfnamefont {J.}~\bibnamefont
  {Casalderrey-Solana}}, \bibinfo {author} {\bibfnamefont {Z.}~\bibnamefont
  {Hulcher}}, \bibinfo {author} {\bibfnamefont {G.}~\bibnamefont {Milhano}},
  \bibinfo {author} {\bibfnamefont {D.}~\bibnamefont {Pablos}}, \ and\ \bibinfo
  {author} {\bibfnamefont {K.}~\bibnamefont {Rajagopal}},\ }\href {\doibase
  10.1103/PhysRevC.99.051901} {\bibfield  {journal} {\bibinfo  {journal} {Phys.
  Rev. C}\ }\textbf {\bibinfo {volume} {99}},\ \bibinfo {pages} {051901}
  (\bibinfo {year} {2019})},\ \Eprint {http://arxiv.org/abs/1808.07386}
  {arXiv:1808.07386 [hep-ph]} \BibitemShut {NoStop}%
\bibitem [{\citenamefont {Brewer}\ \emph {et~al.}(2019)\citenamefont {Brewer},
  \citenamefont {Milhano},\ and\ \citenamefont {Thaler}}]{Brewer:2018dfs}%
  \BibitemOpen
  \bibfield  {author} {\bibinfo {author} {\bibfnamefont {J.}~\bibnamefont
  {Brewer}}, \bibinfo {author} {\bibfnamefont {J.~G.}\ \bibnamefont {Milhano}},
  \ and\ \bibinfo {author} {\bibfnamefont {J.}~\bibnamefont {Thaler}},\ }\href
  {\doibase 10.1103/PhysRevLett.122.222301} {\bibfield  {journal} {\bibinfo
  {journal} {Phys. Rev. Lett.}\ }\textbf {\bibinfo {volume} {122}},\ \bibinfo
  {pages} {222301} (\bibinfo {year} {2019})},\ \Eprint
  {http://arxiv.org/abs/1812.05111} {arXiv:1812.05111 [hep-ph]} \BibitemShut
  {NoStop}%
\bibitem [{\citenamefont {Casalderrey-Solana}\ \emph
  {et~al.}(2020)\citenamefont {Casalderrey-Solana}, \citenamefont {Milhano},
  \citenamefont {Pablos},\ and\ \citenamefont
  {Rajagopal}}]{Casalderrey-Solana:2019ubu}%
  \BibitemOpen
  \bibfield  {author} {\bibinfo {author} {\bibfnamefont {J.}~\bibnamefont
  {Casalderrey-Solana}}, \bibinfo {author} {\bibfnamefont {G.}~\bibnamefont
  {Milhano}}, \bibinfo {author} {\bibfnamefont {D.}~\bibnamefont {Pablos}}, \
  and\ \bibinfo {author} {\bibfnamefont {K.}~\bibnamefont {Rajagopal}},\ }\href
  {\doibase 10.1007/JHEP01(2020)044} {\bibfield  {journal} {\bibinfo  {journal}
  {JHEP}\ }\textbf {\bibinfo {volume} {01}},\ \bibinfo {pages} {044} (\bibinfo
  {year} {2020})},\ \Eprint {http://arxiv.org/abs/1907.11248} {arXiv:1907.11248
  [hep-ph]} \BibitemShut {NoStop}%
\bibitem [{\citenamefont {Caucal}\ \emph {et~al.}(2020)\citenamefont {Caucal},
  \citenamefont {Iancu}, \citenamefont {Mueller},\ and\ \citenamefont
  {Soyez}}]{Caucal:2020xad}%
  \BibitemOpen
  \bibfield  {author} {\bibinfo {author} {\bibfnamefont {P.}~\bibnamefont
  {Caucal}}, \bibinfo {author} {\bibfnamefont {E.}~\bibnamefont {Iancu}},
  \bibinfo {author} {\bibfnamefont {A.~H.}\ \bibnamefont {Mueller}}, \ and\
  \bibinfo {author} {\bibfnamefont {G.}~\bibnamefont {Soyez}},\ }\href
  {\doibase 10.1007/JHEP10(2020)204} {\bibfield  {journal} {\bibinfo  {journal}
  {JHEP}\ }\textbf {\bibinfo {volume} {10}},\ \bibinfo {pages} {204} (\bibinfo
  {year} {2020})},\ \Eprint {http://arxiv.org/abs/2005.05852} {arXiv:2005.05852
  [hep-ph]} \BibitemShut {NoStop}%
\bibitem [{\citenamefont {Du}\ \emph {et~al.}(2020)\citenamefont {Du},
  \citenamefont {Pablos},\ and\ \citenamefont {Tywoniuk}}]{Du:2020pmp}%
  \BibitemOpen
  \bibfield  {author} {\bibinfo {author} {\bibfnamefont {Y.-L.}\ \bibnamefont
  {Du}}, \bibinfo {author} {\bibfnamefont {D.}~\bibnamefont {Pablos}}, \ and\
  \bibinfo {author} {\bibfnamefont {K.}~\bibnamefont {Tywoniuk}},\ }\href
  {\doibase 10.1007/JHEP03(2021)206} {\bibfield  {journal} {\bibinfo  {journal}
  {JHEP}\ }\textbf {\bibinfo {volume} {21}},\ \bibinfo {pages} {206} (\bibinfo
  {year} {2020})},\ \Eprint {http://arxiv.org/abs/2012.07797} {arXiv:2012.07797
  [hep-ph]} \BibitemShut {NoStop}%
\bibitem [{\citenamefont {Du}\ \emph {et~al.}(2022)\citenamefont {Du},
  \citenamefont {Pablos},\ and\ \citenamefont {Tywoniuk}}]{Du:2021pqa}%
  \BibitemOpen
  \bibfield  {author} {\bibinfo {author} {\bibfnamefont {Y.-L.}\ \bibnamefont
  {Du}}, \bibinfo {author} {\bibfnamefont {D.}~\bibnamefont {Pablos}}, \ and\
  \bibinfo {author} {\bibfnamefont {K.}~\bibnamefont {Tywoniuk}},\ }\href
  {\doibase 10.1103/PhysRevLett.128.012301} {\bibfield  {journal} {\bibinfo
  {journal} {Phys. Rev. Lett.}\ }\textbf {\bibinfo {volume} {128}},\ \bibinfo
  {pages} {012301} (\bibinfo {year} {2022})},\ \Eprint
  {http://arxiv.org/abs/2106.11271} {arXiv:2106.11271 [hep-ph]} \BibitemShut
  {NoStop}%
\bibitem [{\citenamefont {Apolin\'ario}\ \emph {et~al.}(2021)\citenamefont
  {Apolin\'ario}, \citenamefont {Castro}, \citenamefont {Crispim Rom\~ao},
  \citenamefont {Milhano}, \citenamefont {Pedro},\ and\ \citenamefont
  {Peres}}]{Apolinario:2021olp}%
  \BibitemOpen
  \bibfield  {author} {\bibinfo {author} {\bibfnamefont {L.}~\bibnamefont
  {Apolin\'ario}}, \bibinfo {author} {\bibfnamefont {N.~F.}\ \bibnamefont
  {Castro}}, \bibinfo {author} {\bibfnamefont {M.}~\bibnamefont {Crispim
  Rom\~ao}}, \bibinfo {author} {\bibfnamefont {J.~G.}\ \bibnamefont {Milhano}},
  \bibinfo {author} {\bibfnamefont {R.}~\bibnamefont {Pedro}}, \ and\ \bibinfo
  {author} {\bibfnamefont {F.~C.~R.}\ \bibnamefont {Peres}},\ }\href {\doibase
  10.1007/JHEP11(2021)219} {\bibfield  {journal} {\bibinfo  {journal} {JHEP}\
  }\textbf {\bibinfo {volume} {11}},\ \bibinfo {pages} {219} (\bibinfo {year}
  {2021})},\ \Eprint {http://arxiv.org/abs/2106.08869} {arXiv:2106.08869
  [hep-ph]} \BibitemShut {NoStop}%
\bibitem [{\citenamefont {Takacs}\ and\ \citenamefont
  {Tywoniuk}(2021)}]{Takacs:2021bpv}%
  \BibitemOpen
  \bibfield  {author} {\bibinfo {author} {\bibfnamefont {A.}~\bibnamefont
  {Takacs}}\ and\ \bibinfo {author} {\bibfnamefont {K.}~\bibnamefont
  {Tywoniuk}},\ }\href {\doibase 10.1007/JHEP10(2021)038} {\bibfield  {journal}
  {\bibinfo  {journal} {JHEP}\ }\textbf {\bibinfo {volume} {10}},\ \bibinfo
  {pages} {038} (\bibinfo {year} {2021})},\ \Eprint
  {http://arxiv.org/abs/2103.14676} {arXiv:2103.14676 [hep-ph]} \BibitemShut
  {NoStop}%
\bibitem [{\citenamefont {Brewer}\ \emph {et~al.}(2022)\citenamefont {Brewer},
  \citenamefont {Brodsky},\ and\ \citenamefont {Rajagopal}}]{Brewer:2021hmh}%
  \BibitemOpen
  \bibfield  {author} {\bibinfo {author} {\bibfnamefont {J.}~\bibnamefont
  {Brewer}}, \bibinfo {author} {\bibfnamefont {Q.}~\bibnamefont {Brodsky}}, \
  and\ \bibinfo {author} {\bibfnamefont {K.}~\bibnamefont {Rajagopal}},\ }\href
  {\doibase 10.1007/JHEP02(2022)175} {\bibfield  {journal} {\bibinfo  {journal}
  {JHEP}\ }\textbf {\bibinfo {volume} {02}},\ \bibinfo {pages} {175} (\bibinfo
  {year} {2022})},\ \Eprint {http://arxiv.org/abs/2110.13159} {arXiv:2110.13159
  [hep-ph]} \BibitemShut {NoStop}%
\bibitem [{\citenamefont {Hayrapetyan}\ \emph
  {et~al.}(2024{\natexlab{c}})\citenamefont {Hayrapetyan} \emph
  {et~al.}}]{CMS:2024zjn}%
  \BibitemOpen
  \bibfield  {author} {\bibinfo {author} {\bibfnamefont {A.}~\bibnamefont
  {Hayrapetyan}} \emph {et~al.} (\bibinfo {collaboration} {CMS}),\ }\href@noop
  {} {\  (\bibinfo {year} {2024}{\natexlab{c}})},\ \Eprint
  {http://arxiv.org/abs/2405.02737} {arXiv:2405.02737 [nucl-ex]} \BibitemShut
  {NoStop}%
\bibitem [{\citenamefont {Sirunyan}\ \emph {et~al.}(2017)\citenamefont
  {Sirunyan} \emph {et~al.}}]{CMS:2017eqd}%
  \BibitemOpen
  \bibfield  {author} {\bibinfo {author} {\bibfnamefont {A.~M.}\ \bibnamefont
  {Sirunyan}} \emph {et~al.} (\bibinfo {collaboration} {CMS}),\ }\href
  {\doibase 10.1103/PhysRevLett.119.082301} {\bibfield  {journal} {\bibinfo
  {journal} {Phys. Rev. Lett.}\ }\textbf {\bibinfo {volume} {119}},\ \bibinfo
  {pages} {082301} (\bibinfo {year} {2017})},\ \Eprint
  {http://arxiv.org/abs/1702.01060} {arXiv:1702.01060 [nucl-ex]} \BibitemShut
  {NoStop}%
\bibitem [{\citenamefont {Sirunyan}\ \emph {et~al.}(2018)\citenamefont
  {Sirunyan} \emph {et~al.}}]{CMS:2017ehl}%
  \BibitemOpen
  \bibfield  {author} {\bibinfo {author} {\bibfnamefont {A.~M.}\ \bibnamefont
  {Sirunyan}} \emph {et~al.} (\bibinfo {collaboration} {CMS}),\ }\href
  {\doibase 10.1016/j.physletb.2018.07.061} {\bibfield  {journal} {\bibinfo
  {journal} {Phys. Lett. B}\ }\textbf {\bibinfo {volume} {785}},\ \bibinfo
  {pages} {14} (\bibinfo {year} {2018})},\ \Eprint
  {http://arxiv.org/abs/1711.09738} {arXiv:1711.09738 [nucl-ex]} \BibitemShut
  {NoStop}%
\bibitem [{\citenamefont {Aaboud}\ \emph {et~al.}(2019)\citenamefont {Aaboud}
  \emph {et~al.}}]{ATLAS:2018dgb}%
  \BibitemOpen
  \bibfield  {author} {\bibinfo {author} {\bibfnamefont {M.}~\bibnamefont
  {Aaboud}} \emph {et~al.} (\bibinfo {collaboration} {ATLAS}),\ }\href
  {\doibase 10.1016/j.physletb.2018.12.023} {\bibfield  {journal} {\bibinfo
  {journal} {Phys. Lett. B}\ }\textbf {\bibinfo {volume} {789}},\ \bibinfo
  {pages} {167} (\bibinfo {year} {2019})},\ \Eprint
  {http://arxiv.org/abs/1809.07280} {arXiv:1809.07280 [nucl-ex]} \BibitemShut
  {NoStop}%
\bibitem [{\citenamefont {Sirunyan}\ \emph {et~al.}(2019)\citenamefont
  {Sirunyan} \emph {et~al.}}]{CMS:2018jco}%
  \BibitemOpen
  \bibfield  {author} {\bibinfo {author} {\bibfnamefont {A.~M.}\ \bibnamefont
  {Sirunyan}} \emph {et~al.} (\bibinfo {collaboration} {CMS}),\ }\href
  {\doibase 10.1103/PhysRevLett.122.152001} {\bibfield  {journal} {\bibinfo
  {journal} {Phys. Rev. Lett.}\ }\textbf {\bibinfo {volume} {122}},\ \bibinfo
  {pages} {152001} (\bibinfo {year} {2019})},\ \Eprint
  {http://arxiv.org/abs/1809.08602} {arXiv:1809.08602 [hep-ex]} \BibitemShut
  {NoStop}%
\bibitem [{\citenamefont {Aad}\ \emph {et~al.}(2021)\citenamefont {Aad} \emph
  {et~al.}}]{ATLAS:2020wmg}%
  \BibitemOpen
  \bibfield  {author} {\bibinfo {author} {\bibfnamefont {G.}~\bibnamefont
  {Aad}} \emph {et~al.} (\bibinfo {collaboration} {ATLAS}),\ }\href {\doibase
  10.1103/PhysRevLett.126.072301} {\bibfield  {journal} {\bibinfo  {journal}
  {Phys. Rev. Lett.}\ }\textbf {\bibinfo {volume} {126}},\ \bibinfo {pages}
  {072301} (\bibinfo {year} {2021})},\ \Eprint
  {http://arxiv.org/abs/2008.09811} {arXiv:2008.09811 [nucl-ex]} \BibitemShut
  {NoStop}%
\bibitem [{\citenamefont {Chen}\ \emph {et~al.}(2020)\citenamefont {Chen},
  \citenamefont {Moult}, \citenamefont {Zhang},\ and\ \citenamefont
  {Zhu}}]{Chen:2020vvp}%
  \BibitemOpen
  \bibfield  {author} {\bibinfo {author} {\bibfnamefont {H.}~\bibnamefont
  {Chen}}, \bibinfo {author} {\bibfnamefont {I.}~\bibnamefont {Moult}},
  \bibinfo {author} {\bibfnamefont {X.}~\bibnamefont {Zhang}}, \ and\ \bibinfo
  {author} {\bibfnamefont {H.~X.}\ \bibnamefont {Zhu}},\ }\href {\doibase
  10.1103/PhysRevD.102.054012} {\bibfield  {journal} {\bibinfo  {journal}
  {Phys. Rev. D}\ }\textbf {\bibinfo {volume} {102}},\ \bibinfo {pages}
  {054012} (\bibinfo {year} {2020})},\ \Eprint
  {http://arxiv.org/abs/2004.11381} {arXiv:2004.11381 [hep-ph]} \BibitemShut
  {NoStop}%
\bibitem [{\citenamefont {Komiske}\ \emph {et~al.}(2023)\citenamefont
  {Komiske}, \citenamefont {Moult}, \citenamefont {Thaler},\ and\ \citenamefont
  {Zhu}}]{Komiske:2022enw}%
  \BibitemOpen
  \bibfield  {author} {\bibinfo {author} {\bibfnamefont {P.~T.}\ \bibnamefont
  {Komiske}}, \bibinfo {author} {\bibfnamefont {I.}~\bibnamefont {Moult}},
  \bibinfo {author} {\bibfnamefont {J.}~\bibnamefont {Thaler}}, \ and\ \bibinfo
  {author} {\bibfnamefont {H.~X.}\ \bibnamefont {Zhu}},\ }\href {\doibase
  10.1103/PhysRevLett.130.051901} {\bibfield  {journal} {\bibinfo  {journal}
  {Phys. Rev. Lett.}\ }\textbf {\bibinfo {volume} {130}},\ \bibinfo {pages}
  {051901} (\bibinfo {year} {2023})},\ \Eprint
  {http://arxiv.org/abs/2201.07800} {arXiv:2201.07800 [hep-ph]} \BibitemShut
  {NoStop}%
\bibitem [{\citenamefont {Lee}\ \emph {et~al.}(2022)\citenamefont {Lee},
  \citenamefont {Me\c{c}aj},\ and\ \citenamefont {Moult}}]{Lee:2022ige}%
  \BibitemOpen
  \bibfield  {author} {\bibinfo {author} {\bibfnamefont {K.}~\bibnamefont
  {Lee}}, \bibinfo {author} {\bibfnamefont {B.}~\bibnamefont {Me\c{c}aj}}, \
  and\ \bibinfo {author} {\bibfnamefont {I.}~\bibnamefont {Moult}},\
  }\href@noop {} {\  (\bibinfo {year} {2022})},\ \Eprint
  {http://arxiv.org/abs/2205.03414} {arXiv:2205.03414 [hep-ph]} \BibitemShut
  {NoStop}%
\bibitem [{\citenamefont {Sveshnikov}\ and\ \citenamefont
  {Tkachov}(1996)}]{Sveshnikov:1995vi}%
  \BibitemOpen
  \bibfield  {author} {\bibinfo {author} {\bibfnamefont {N.}~\bibnamefont
  {Sveshnikov}}\ and\ \bibinfo {author} {\bibfnamefont {F.}~\bibnamefont
  {Tkachov}},\ }\href {\doibase 10.1016/0370-2693(96)00558-8} {\bibfield
  {journal} {\bibinfo  {journal} {Phys. Lett. B}\ }\textbf {\bibinfo {volume}
  {382}},\ \bibinfo {pages} {403} (\bibinfo {year} {1996})},\ \Eprint
  {http://arxiv.org/abs/hep-ph/9512370} {arXiv:hep-ph/9512370} \BibitemShut
  {NoStop}%
\bibitem [{\citenamefont {Tkachov}(1997)}]{Tkachov:1995kk}%
  \BibitemOpen
  \bibfield  {author} {\bibinfo {author} {\bibfnamefont {F.~V.}\ \bibnamefont
  {Tkachov}},\ }\href {\doibase 10.1142/S0217751X97002899} {\bibfield
  {journal} {\bibinfo  {journal} {Int. J. Mod. Phys. A}\ }\textbf {\bibinfo
  {volume} {12}},\ \bibinfo {pages} {5411} (\bibinfo {year} {1997})},\ \Eprint
  {http://arxiv.org/abs/hep-ph/9601308} {arXiv:hep-ph/9601308} \BibitemShut
  {NoStop}%
\bibitem [{\citenamefont {Korchemsky}\ and\ \citenamefont
  {Sterman}(1999)}]{Korchemsky:1999kt}%
  \BibitemOpen
  \bibfield  {author} {\bibinfo {author} {\bibfnamefont {G.~P.}\ \bibnamefont
  {Korchemsky}}\ and\ \bibinfo {author} {\bibfnamefont {G.~F.}\ \bibnamefont
  {Sterman}},\ }\href {\doibase 10.1016/S0550-3213(99)00308-9} {\bibfield
  {journal} {\bibinfo  {journal} {Nucl. Phys. B}\ }\textbf {\bibinfo {volume}
  {555}},\ \bibinfo {pages} {335} (\bibinfo {year} {1999})},\ \Eprint
  {http://arxiv.org/abs/hep-ph/9902341} {arXiv:hep-ph/9902341} \BibitemShut
  {NoStop}%
\bibitem [{\citenamefont {Bauer}\ \emph {et~al.}(2008)\citenamefont {Bauer},
  \citenamefont {Fleming}, \citenamefont {Lee},\ and\ \citenamefont
  {Sterman}}]{Bauer:2008dt}%
  \BibitemOpen
  \bibfield  {author} {\bibinfo {author} {\bibfnamefont {C.~W.}\ \bibnamefont
  {Bauer}}, \bibinfo {author} {\bibfnamefont {S.~P.}\ \bibnamefont {Fleming}},
  \bibinfo {author} {\bibfnamefont {C.}~\bibnamefont {Lee}}, \ and\ \bibinfo
  {author} {\bibfnamefont {G.~F.}\ \bibnamefont {Sterman}},\ }\href {\doibase
  10.1103/PhysRevD.78.034027} {\bibfield  {journal} {\bibinfo  {journal} {Phys.
  Rev. D}\ }\textbf {\bibinfo {volume} {78}},\ \bibinfo {pages} {034027}
  (\bibinfo {year} {2008})},\ \Eprint {http://arxiv.org/abs/0801.4569}
  {arXiv:0801.4569 [hep-ph]} \BibitemShut {NoStop}%
\bibitem [{\citenamefont {Belitsky}\ \emph
  {et~al.}(2014{\natexlab{a}})\citenamefont {Belitsky}, \citenamefont
  {Hohenegger}, \citenamefont {Korchemsky}, \citenamefont {Sokatchev},\ and\
  \citenamefont {Zhiboedov}}]{Belitsky:2013xxa}%
  \BibitemOpen
  \bibfield  {author} {\bibinfo {author} {\bibfnamefont {A.}~\bibnamefont
  {Belitsky}}, \bibinfo {author} {\bibfnamefont {S.}~\bibnamefont
  {Hohenegger}}, \bibinfo {author} {\bibfnamefont {G.}~\bibnamefont
  {Korchemsky}}, \bibinfo {author} {\bibfnamefont {E.}~\bibnamefont
  {Sokatchev}}, \ and\ \bibinfo {author} {\bibfnamefont {A.}~\bibnamefont
  {Zhiboedov}},\ }\href {\doibase 10.1016/j.nuclphysb.2014.04.020} {\bibfield
  {journal} {\bibinfo  {journal} {Nucl. Phys. B}\ }\textbf {\bibinfo {volume}
  {884}},\ \bibinfo {pages} {305} (\bibinfo {year} {2014}{\natexlab{a}})},\
  \Eprint {http://arxiv.org/abs/1309.0769} {arXiv:1309.0769 [hep-th]}
  \BibitemShut {NoStop}%
\bibitem [{\citenamefont {Belitsky}\ \emph
  {et~al.}(2014{\natexlab{b}})\citenamefont {Belitsky}, \citenamefont
  {Hohenegger}, \citenamefont {Korchemsky}, \citenamefont {Sokatchev},\ and\
  \citenamefont {Zhiboedov}}]{Belitsky:2013bja}%
  \BibitemOpen
  \bibfield  {author} {\bibinfo {author} {\bibfnamefont {A.}~\bibnamefont
  {Belitsky}}, \bibinfo {author} {\bibfnamefont {S.}~\bibnamefont
  {Hohenegger}}, \bibinfo {author} {\bibfnamefont {G.}~\bibnamefont
  {Korchemsky}}, \bibinfo {author} {\bibfnamefont {E.}~\bibnamefont
  {Sokatchev}}, \ and\ \bibinfo {author} {\bibfnamefont {A.}~\bibnamefont
  {Zhiboedov}},\ }\href {\doibase 10.1016/j.nuclphysb.2014.04.019} {\bibfield
  {journal} {\bibinfo  {journal} {Nucl. Phys. B}\ }\textbf {\bibinfo {volume}
  {884}},\ \bibinfo {pages} {206} (\bibinfo {year} {2014}{\natexlab{b}})},\
  \Eprint {http://arxiv.org/abs/1309.1424} {arXiv:1309.1424 [hep-th]}
  \BibitemShut {NoStop}%
\bibitem [{\citenamefont {Kravchuk}\ and\ \citenamefont
  {Simmons-Duffin}(2018)}]{Kravchuk:2018htv}%
  \BibitemOpen
  \bibfield  {author} {\bibinfo {author} {\bibfnamefont {P.}~\bibnamefont
  {Kravchuk}}\ and\ \bibinfo {author} {\bibfnamefont {D.}~\bibnamefont
  {Simmons-Duffin}},\ }\href {\doibase 10.1007/JHEP11(2018)102} {\bibfield
  {journal} {\bibinfo  {journal} {JHEP}\ }\textbf {\bibinfo {volume} {11}},\
  \bibinfo {pages} {102} (\bibinfo {year} {2018})},\ \Eprint
  {http://arxiv.org/abs/1805.00098} {arXiv:1805.00098 [hep-th]} \BibitemShut
  {NoStop}%
\bibitem [{\citenamefont {Fan}({\natexlab{b}})}]{talk1}%
  \BibitemOpen
  \bibfield  {author} {\bibinfo {author} {\bibfnamefont {W.}~\bibnamefont
  {Fan}},\ }\href@noop {} {\bibfield  {journal} {\bibinfo  {journal} {Imaging
  Cold Nuclear Matter with Energy Correlators at the future EIC, DIS 2023}\ }
  ({\natexlab{b}})}\BibitemShut {NoStop}%
\bibitem [{\citenamefont {Salgado}\ and\ \citenamefont
  {Wiedemann}(2003)}]{Salgado:2003gb}%
  \BibitemOpen
  \bibfield  {author} {\bibinfo {author} {\bibfnamefont {C.~A.}\ \bibnamefont
  {Salgado}}\ and\ \bibinfo {author} {\bibfnamefont {U.~A.}\ \bibnamefont
  {Wiedemann}},\ }\href {\doibase 10.1103/PhysRevD.68.014008} {\bibfield
  {journal} {\bibinfo  {journal} {Phys. Rev. D}\ }\textbf {\bibinfo {volume}
  {68}},\ \bibinfo {pages} {014008} (\bibinfo {year} {2003})},\ \Eprint
  {http://arxiv.org/abs/hep-ph/0302184} {arXiv:hep-ph/0302184} \BibitemShut
  {NoStop}%
\bibitem [{\citenamefont {Mehtar-Tani}\ \emph {et~al.}(2023)\citenamefont
  {Mehtar-Tani}, \citenamefont {Schlichting},\ and\ \citenamefont
  {Soudi}}]{Mehtar-Tani:2022zwf}%
  \BibitemOpen
  \bibfield  {author} {\bibinfo {author} {\bibfnamefont {Y.}~\bibnamefont
  {Mehtar-Tani}}, \bibinfo {author} {\bibfnamefont {S.}~\bibnamefont
  {Schlichting}}, \ and\ \bibinfo {author} {\bibfnamefont {I.}~\bibnamefont
  {Soudi}},\ }\href {\doibase 10.1007/JHEP05(2023)091} {\bibfield  {journal}
  {\bibinfo  {journal} {JHEP}\ }\textbf {\bibinfo {volume} {05}},\ \bibinfo
  {pages} {091} (\bibinfo {year} {2023})},\ \Eprint
  {http://arxiv.org/abs/2209.10569} {arXiv:2209.10569 [hep-ph]} \BibitemShut
  {NoStop}%
\bibitem [{\citenamefont {Kologlu}\ \emph {et~al.}(2021)\citenamefont
  {Kologlu}, \citenamefont {Kravchuk}, \citenamefont {Simmons-Duffin},\ and\
  \citenamefont {Zhiboedov}}]{Kologlu:2019mfz}%
  \BibitemOpen
  \bibfield  {author} {\bibinfo {author} {\bibfnamefont {M.}~\bibnamefont
  {Kologlu}}, \bibinfo {author} {\bibfnamefont {P.}~\bibnamefont {Kravchuk}},
  \bibinfo {author} {\bibfnamefont {D.}~\bibnamefont {Simmons-Duffin}}, \ and\
  \bibinfo {author} {\bibfnamefont {A.}~\bibnamefont {Zhiboedov}},\ }\href
  {\doibase 10.1007/JHEP01(2021)128} {\bibfield  {journal} {\bibinfo  {journal}
  {JHEP}\ }\textbf {\bibinfo {volume} {01}},\ \bibinfo {pages} {128} (\bibinfo
  {year} {2021})},\ \Eprint {http://arxiv.org/abs/1905.01311} {arXiv:1905.01311
  [hep-th]} \BibitemShut {NoStop}%
\bibitem [{\citenamefont {Dixon}\ \emph {et~al.}(2019)\citenamefont {Dixon},
  \citenamefont {Moult},\ and\ \citenamefont {Zhu}}]{Dixon:2019uzg}%
  \BibitemOpen
  \bibfield  {author} {\bibinfo {author} {\bibfnamefont {L.~J.}\ \bibnamefont
  {Dixon}}, \bibinfo {author} {\bibfnamefont {I.}~\bibnamefont {Moult}}, \ and\
  \bibinfo {author} {\bibfnamefont {H.~X.}\ \bibnamefont {Zhu}},\ }\href
  {\doibase 10.1103/PhysRevD.100.014009} {\bibfield  {journal} {\bibinfo
  {journal} {Phys. Rev. D}\ }\textbf {\bibinfo {volume} {100}},\ \bibinfo
  {pages} {014009} (\bibinfo {year} {2019})},\ \Eprint
  {http://arxiv.org/abs/1905.01310} {arXiv:1905.01310 [hep-ph]} \BibitemShut
  {NoStop}%
\bibitem [{\citenamefont {Schindler}\ \emph {et~al.}(2023)\citenamefont
  {Schindler}, \citenamefont {Stewart},\ and\ \citenamefont
  {Sun}}]{Schindler:2023cww}%
  \BibitemOpen
  \bibfield  {author} {\bibinfo {author} {\bibfnamefont {S.~T.}\ \bibnamefont
  {Schindler}}, \bibinfo {author} {\bibfnamefont {I.~W.}\ \bibnamefont
  {Stewart}}, \ and\ \bibinfo {author} {\bibfnamefont {Z.}~\bibnamefont
  {Sun}},\ }\href {\doibase 10.1007/JHEP10(2023)187} {\bibfield  {journal}
  {\bibinfo  {journal} {JHEP}\ }\textbf {\bibinfo {volume} {10}},\ \bibinfo
  {pages} {187} (\bibinfo {year} {2023})},\ \Eprint
  {http://arxiv.org/abs/2305.19311} {arXiv:2305.19311 [hep-ph]} \BibitemShut
  {NoStop}%
\bibitem [{\citenamefont {Lee}\ \emph {et~al.}(2024)\citenamefont {Lee},
  \citenamefont {Pathak}, \citenamefont {Stewart},\ and\ \citenamefont
  {Sun}}]{Lee:2024esz}%
  \BibitemOpen
  \bibfield  {author} {\bibinfo {author} {\bibfnamefont {K.}~\bibnamefont
  {Lee}}, \bibinfo {author} {\bibfnamefont {A.}~\bibnamefont {Pathak}},
  \bibinfo {author} {\bibfnamefont {I.}~\bibnamefont {Stewart}}, \ and\
  \bibinfo {author} {\bibfnamefont {Z.}~\bibnamefont {Sun}},\ }\href@noop {} {\
   (\bibinfo {year} {2024})},\ \Eprint {http://arxiv.org/abs/2405.19396}
  {arXiv:2405.19396 [hep-ph]} \BibitemShut {NoStop}%
\bibitem [{\citenamefont {Chen}\ \emph {et~al.}(2024)\citenamefont {Chen},
  \citenamefont {Monni}, \citenamefont {Xu},\ and\ \citenamefont
  {Zhu}}]{Chen:2024nyc}%
  \BibitemOpen
  \bibfield  {author} {\bibinfo {author} {\bibfnamefont {H.}~\bibnamefont
  {Chen}}, \bibinfo {author} {\bibfnamefont {P.~F.}\ \bibnamefont {Monni}},
  \bibinfo {author} {\bibfnamefont {Z.}~\bibnamefont {Xu}}, \ and\ \bibinfo
  {author} {\bibfnamefont {H.~X.}\ \bibnamefont {Zhu}},\ }\href@noop {} {\
  (\bibinfo {year} {2024})},\ \Eprint {http://arxiv.org/abs/2406.06668}
  {arXiv:2406.06668 [hep-ph]} \BibitemShut {NoStop}%
\bibitem [{\citenamefont {Sj\"ostrand}\ \emph {et~al.}(2015)\citenamefont
  {Sj\"ostrand}, \citenamefont {Ask}, \citenamefont {Christiansen},
  \citenamefont {Corke}, \citenamefont {Desai}, \citenamefont {Ilten},
  \citenamefont {Mrenna}, \citenamefont {Prestel}, \citenamefont {Rasmussen},\
  and\ \citenamefont {Skands}}]{Sjostrand:2014zea}%
  \BibitemOpen
  \bibfield  {author} {\bibinfo {author} {\bibfnamefont {T.}~\bibnamefont
  {Sj\"ostrand}}, \bibinfo {author} {\bibfnamefont {S.}~\bibnamefont {Ask}},
  \bibinfo {author} {\bibfnamefont {J.~R.}\ \bibnamefont {Christiansen}},
  \bibinfo {author} {\bibfnamefont {R.}~\bibnamefont {Corke}}, \bibinfo
  {author} {\bibfnamefont {N.}~\bibnamefont {Desai}}, \bibinfo {author}
  {\bibfnamefont {P.}~\bibnamefont {Ilten}}, \bibinfo {author} {\bibfnamefont
  {S.}~\bibnamefont {Mrenna}}, \bibinfo {author} {\bibfnamefont
  {S.}~\bibnamefont {Prestel}}, \bibinfo {author} {\bibfnamefont {C.~O.}\
  \bibnamefont {Rasmussen}}, \ and\ \bibinfo {author} {\bibfnamefont {P.~Z.}\
  \bibnamefont {Skands}},\ }\href {\doibase 10.1016/j.cpc.2015.01.024}
  {\bibfield  {journal} {\bibinfo  {journal} {Comput. Phys. Commun.}\ }\textbf
  {\bibinfo {volume} {191}},\ \bibinfo {pages} {159} (\bibinfo {year}
  {2015})},\ \Eprint {http://arxiv.org/abs/1410.3012} {arXiv:1410.3012
  [hep-ph]} \BibitemShut {NoStop}%
\bibitem [{\citenamefont {Bellm}\ \emph {et~al.}(2016)\citenamefont {Bellm}
  \emph {et~al.}}]{Bellm:2015jjp}%
  \BibitemOpen
  \bibfield  {author} {\bibinfo {author} {\bibfnamefont {J.}~\bibnamefont
  {Bellm}} \emph {et~al.},\ }\href {\doibase 10.1140/epjc/s10052-016-4018-8}
  {\bibfield  {journal} {\bibinfo  {journal} {Eur. Phys. J. C}\ }\textbf
  {\bibinfo {volume} {76}},\ \bibinfo {pages} {196} (\bibinfo {year} {2016})},\
  \Eprint {http://arxiv.org/abs/1512.01178} {arXiv:1512.01178 [hep-ph]}
  \BibitemShut {NoStop}%
\bibitem [{\citenamefont {Bahr}\ \emph {et~al.}(2008)\citenamefont {Bahr} \emph
  {et~al.}}]{Bahr:2008pv}%
  \BibitemOpen
  \bibfield  {author} {\bibinfo {author} {\bibfnamefont {M.}~\bibnamefont
  {Bahr}} \emph {et~al.},\ }\href {\doibase 10.1140/epjc/s10052-008-0798-9}
  {\bibfield  {journal} {\bibinfo  {journal} {Eur. Phys. J. C}\ }\textbf
  {\bibinfo {volume} {58}},\ \bibinfo {pages} {639} (\bibinfo {year} {2008})},\
  \Eprint {http://arxiv.org/abs/0803.0883} {arXiv:0803.0883 [hep-ph]}
  \BibitemShut {NoStop}%
\bibitem [{\citenamefont {Sirunyan}\ \emph {et~al.}(2020)\citenamefont
  {Sirunyan} \emph {et~al.}}]{CMS:pythia8tune}%
  \BibitemOpen
  \bibfield  {author} {\bibinfo {author} {\bibfnamefont {A.~M.}\ \bibnamefont
  {Sirunyan}} \emph {et~al.} (\bibinfo {collaboration} {CMS}),\ }\href
  {\doibase 10.1140/epjc/s10052-019-7499-4} {\bibfield  {journal} {\bibinfo
  {journal} {Eur. Phys. J. C}\ }\textbf {\bibinfo {volume} {80}},\ \bibinfo
  {pages} {4} (\bibinfo {year} {2020})},\ \Eprint
  {http://arxiv.org/abs/1903.12179} {arXiv:1903.12179 [hep-ex]} \BibitemShut
  {NoStop}%
\bibitem [{\citenamefont {Sirunyan}\ \emph {et~al.}(2021)\citenamefont
  {Sirunyan} \emph {et~al.}}]{CMS:herwig7tune}%
  \BibitemOpen
  \bibfield  {author} {\bibinfo {author} {\bibfnamefont {A.~M.}\ \bibnamefont
  {Sirunyan}} \emph {et~al.} (\bibinfo {collaboration} {CMS}),\ }\href
  {\doibase 10.1140/epjc/s10052-021-08949-5} {\bibfield  {journal} {\bibinfo
  {journal} {Eur. Phys. J. C}\ }\textbf {\bibinfo {volume} {81}},\ \bibinfo
  {pages} {312} (\bibinfo {year} {2021})},\ \Eprint
  {http://arxiv.org/abs/2011.03422} {arXiv:2011.03422 [hep-ex]} \BibitemShut
  {NoStop}%
\bibitem [{\citenamefont {Cacciari}\ \emph {et~al.}(2008)\citenamefont
  {Cacciari}, \citenamefont {Salam},\ and\ \citenamefont
  {Soyez}}]{Cacciari:2008gp}%
  \BibitemOpen
  \bibfield  {author} {\bibinfo {author} {\bibfnamefont {M.}~\bibnamefont
  {Cacciari}}, \bibinfo {author} {\bibfnamefont {G.~P.}\ \bibnamefont {Salam}},
  \ and\ \bibinfo {author} {\bibfnamefont {G.}~\bibnamefont {Soyez}},\ }\href
  {\doibase 10.1088/1126-6708/2008/04/063} {\bibfield  {journal} {\bibinfo
  {journal} {JHEP}\ }\textbf {\bibinfo {volume} {04}},\ \bibinfo {pages} {063}
  (\bibinfo {year} {2008})},\ \Eprint {http://arxiv.org/abs/0802.1189}
  {arXiv:0802.1189 [hep-ph]} \BibitemShut {NoStop}%
\bibitem [{\citenamefont {Andres}\ \emph
  {et~al.}(2023{\natexlab{c}})\citenamefont {Andres}, \citenamefont
  {Dominguez}, \citenamefont {Holguin}, \citenamefont {Marquet},\ and\
  \citenamefont {Moult}}]{Andres:2023ymw}%
  \BibitemOpen
  \bibfield  {author} {\bibinfo {author} {\bibfnamefont {C.}~\bibnamefont
  {Andres}}, \bibinfo {author} {\bibfnamefont {F.}~\bibnamefont {Dominguez}},
  \bibinfo {author} {\bibfnamefont {J.}~\bibnamefont {Holguin}}, \bibinfo
  {author} {\bibfnamefont {C.}~\bibnamefont {Marquet}}, \ and\ \bibinfo
  {author} {\bibfnamefont {I.}~\bibnamefont {Moult}},\ }\href@noop {} {\
  (\bibinfo {year} {2023}{\natexlab{c}})},\ \Eprint
  {http://arxiv.org/abs/2307.15110} {arXiv:2307.15110 [hep-ph]} \BibitemShut
  {NoStop}%
\bibitem [{\citenamefont {Barata}\ \emph {et~al.}(2024)\citenamefont {Barata},
  \citenamefont {Milhano},\ and\ \citenamefont {Sadofyev}}]{Barata:2023zqg}%
  \BibitemOpen
  \bibfield  {author} {\bibinfo {author} {\bibfnamefont {J.~a.}\ \bibnamefont
  {Barata}}, \bibinfo {author} {\bibfnamefont {J.~G.}\ \bibnamefont {Milhano}},
  \ and\ \bibinfo {author} {\bibfnamefont {A.~V.}\ \bibnamefont {Sadofyev}},\
  }\href {\doibase 10.1140/epjc/s10052-024-12514-1} {\bibfield  {journal}
  {\bibinfo  {journal} {Eur. Phys. J. C}\ }\textbf {\bibinfo {volume} {84}},\
  \bibinfo {pages} {174} (\bibinfo {year} {2024})},\ \Eprint
  {http://arxiv.org/abs/2308.01294} {arXiv:2308.01294 [hep-ph]} \BibitemShut
  {NoStop}%
\bibitem [{\citenamefont {Singh}\ and\ \citenamefont
  {Vaidya}(2024)}]{Singh:2024vwb}%
  \BibitemOpen
  \bibfield  {author} {\bibinfo {author} {\bibfnamefont {B.}~\bibnamefont
  {Singh}}\ and\ \bibinfo {author} {\bibfnamefont {V.}~\bibnamefont {Vaidya}},\
  }\href@noop {} {\  (\bibinfo {year} {2024})},\ \Eprint
  {http://arxiv.org/abs/2408.02753} {arXiv:2408.02753 [hep-ph]} \BibitemShut
  {NoStop}%
\bibitem [{\citenamefont {Holguin}\ \emph
  {et~al.}(2023{\natexlab{a}})\citenamefont {Holguin}, \citenamefont {Moult},
  \citenamefont {Pathak},\ and\ \citenamefont {Procura}}]{Holguin:2022epo}%
  \BibitemOpen
  \bibfield  {author} {\bibinfo {author} {\bibfnamefont {J.}~\bibnamefont
  {Holguin}}, \bibinfo {author} {\bibfnamefont {I.}~\bibnamefont {Moult}},
  \bibinfo {author} {\bibfnamefont {A.}~\bibnamefont {Pathak}}, \ and\ \bibinfo
  {author} {\bibfnamefont {M.}~\bibnamefont {Procura}},\ }\href {\doibase
  10.1103/PhysRevD.107.114002} {\bibfield  {journal} {\bibinfo  {journal}
  {Phys. Rev. D}\ }\textbf {\bibinfo {volume} {107}},\ \bibinfo {pages}
  {114002} (\bibinfo {year} {2023}{\natexlab{a}})},\ \Eprint
  {http://arxiv.org/abs/2201.08393} {arXiv:2201.08393 [hep-ph]} \BibitemShut
  {NoStop}%
\bibitem [{\citenamefont {Holguin}\ \emph
  {et~al.}(2023{\natexlab{b}})\citenamefont {Holguin}, \citenamefont {Moult},
  \citenamefont {Pathak}, \citenamefont {Procura}, \citenamefont
  {Sch\"ofbeck},\ and\ \citenamefont {Schwarz}}]{Holguin:2023bjf}%
  \BibitemOpen
  \bibfield  {author} {\bibinfo {author} {\bibfnamefont {J.}~\bibnamefont
  {Holguin}}, \bibinfo {author} {\bibfnamefont {I.}~\bibnamefont {Moult}},
  \bibinfo {author} {\bibfnamefont {A.}~\bibnamefont {Pathak}}, \bibinfo
  {author} {\bibfnamefont {M.}~\bibnamefont {Procura}}, \bibinfo {author}
  {\bibfnamefont {R.}~\bibnamefont {Sch\"ofbeck}}, \ and\ \bibinfo {author}
  {\bibfnamefont {D.}~\bibnamefont {Schwarz}},\ }\href@noop {} {\  (\bibinfo
  {year} {2023}{\natexlab{b}})},\ \Eprint {http://arxiv.org/abs/2311.02157}
  {arXiv:2311.02157 [hep-ph]} \BibitemShut {NoStop}%
\bibitem [{\citenamefont {Holguin}\ \emph {et~al.}(2024)\citenamefont
  {Holguin}, \citenamefont {Moult}, \citenamefont {Pathak}, \citenamefont
  {Procura}, \citenamefont {Sch\"ofbeck},\ and\ \citenamefont
  {Schwarz}}]{Holguin:2024tkz}%
  \BibitemOpen
  \bibfield  {author} {\bibinfo {author} {\bibfnamefont {J.}~\bibnamefont
  {Holguin}}, \bibinfo {author} {\bibfnamefont {I.}~\bibnamefont {Moult}},
  \bibinfo {author} {\bibfnamefont {A.}~\bibnamefont {Pathak}}, \bibinfo
  {author} {\bibfnamefont {M.}~\bibnamefont {Procura}}, \bibinfo {author}
  {\bibfnamefont {R.}~\bibnamefont {Sch\"ofbeck}}, \ and\ \bibinfo {author}
  {\bibfnamefont {D.}~\bibnamefont {Schwarz}},\ }\href@noop {} {\  (\bibinfo
  {year} {2024})},\ \Eprint {http://arxiv.org/abs/2407.12900} {arXiv:2407.12900
  [hep-ph]} \BibitemShut {NoStop}%
\bibitem [{\citenamefont {Liu}\ and\ \citenamefont {Zhu}(2023)}]{Liu:2022wop}%
  \BibitemOpen
  \bibfield  {author} {\bibinfo {author} {\bibfnamefont {X.}~\bibnamefont
  {Liu}}\ and\ \bibinfo {author} {\bibfnamefont {H.~X.}\ \bibnamefont {Zhu}},\
  }\href {\doibase 10.1103/PhysRevLett.130.091901} {\bibfield  {journal}
  {\bibinfo  {journal} {Phys. Rev. Lett.}\ }\textbf {\bibinfo {volume} {130}},\
  \bibinfo {pages} {091901} (\bibinfo {year} {2023})},\ \Eprint
  {http://arxiv.org/abs/2209.02080} {arXiv:2209.02080 [hep-ph]} \BibitemShut
  {NoStop}%
\bibitem [{\citenamefont {Liu}\ \emph {et~al.}(2023)\citenamefont {Liu},
  \citenamefont {Liu}, \citenamefont {Pan}, \citenamefont {Yuan},\ and\
  \citenamefont {Zhu}}]{Liu:2023aqb}%
  \BibitemOpen
  \bibfield  {author} {\bibinfo {author} {\bibfnamefont {H.-Y.}\ \bibnamefont
  {Liu}}, \bibinfo {author} {\bibfnamefont {X.}~\bibnamefont {Liu}}, \bibinfo
  {author} {\bibfnamefont {J.-C.}\ \bibnamefont {Pan}}, \bibinfo {author}
  {\bibfnamefont {F.}~\bibnamefont {Yuan}}, \ and\ \bibinfo {author}
  {\bibfnamefont {H.~X.}\ \bibnamefont {Zhu}},\ }\href {\doibase
  10.1103/PhysRevLett.130.181901} {\bibfield  {journal} {\bibinfo  {journal}
  {Phys. Rev. Lett.}\ }\textbf {\bibinfo {volume} {130}},\ \bibinfo {pages}
  {181901} (\bibinfo {year} {2023})},\ \Eprint
  {http://arxiv.org/abs/2301.01788} {arXiv:2301.01788 [hep-ph]} \BibitemShut
  {NoStop}%
\bibitem [{\citenamefont {Cao}\ \emph {et~al.}(2023)\citenamefont {Cao},
  \citenamefont {Liu},\ and\ \citenamefont {Zhu}}]{Cao:2023oef}%
  \BibitemOpen
  \bibfield  {author} {\bibinfo {author} {\bibfnamefont {H.}~\bibnamefont
  {Cao}}, \bibinfo {author} {\bibfnamefont {X.}~\bibnamefont {Liu}}, \ and\
  \bibinfo {author} {\bibfnamefont {H.~X.}\ \bibnamefont {Zhu}},\ }\href
  {\doibase 10.1103/PhysRevD.107.114008} {\bibfield  {journal} {\bibinfo
  {journal} {Phys. Rev. D}\ }\textbf {\bibinfo {volume} {107}},\ \bibinfo
  {pages} {114008} (\bibinfo {year} {2023})},\ \Eprint
  {http://arxiv.org/abs/2303.01530} {arXiv:2303.01530 [hep-ph]} \BibitemShut
  {NoStop}%
\end{thebibliography}%
\bibliographystyle{apsrev4-1}
\newpage
\onecolumngrid
\newpage
\newpage

\begin{center}
    \large{\textbf{Supplemental material}}
\end{center}

In this supplemental material, we present additional figures demonstrating the performance of the unbiasing factor, $C_{2}$, across different jet $p_{T}$ ranges. As in the main text, we used inclusive jet samples in $\sqrt{s}=5.02~$TeV p-p collisions generated with both \HERWIG 7.2.2~\cite{Bellm:2015jjp,Bahr:2008pv} and \PYTHIA 8.230~\cite{Sjostrand:2014zea} Monte Carlo event generators. For \HERWIG, around 7 million events were generated with the CH3 tune~\cite{CMS:herwig7tune}, while for \PYTHIA, approximately 44 million events were generated with the CP5 tune~\cite{CMS:pythia8tune}. Inclusive jets were clustered using the anti-$k_T$ algorithm~\cite{Cacciari:2008gp} with a jet radius of $R=0.4$. The E2C distribution was computed for charged particles  following the analysis described in~\cite{talkEEC,CMS-PAS-HIN-23-004}.

\begin{figure}[ht]
    \centering
    \includegraphics[width=0.49\linewidth]{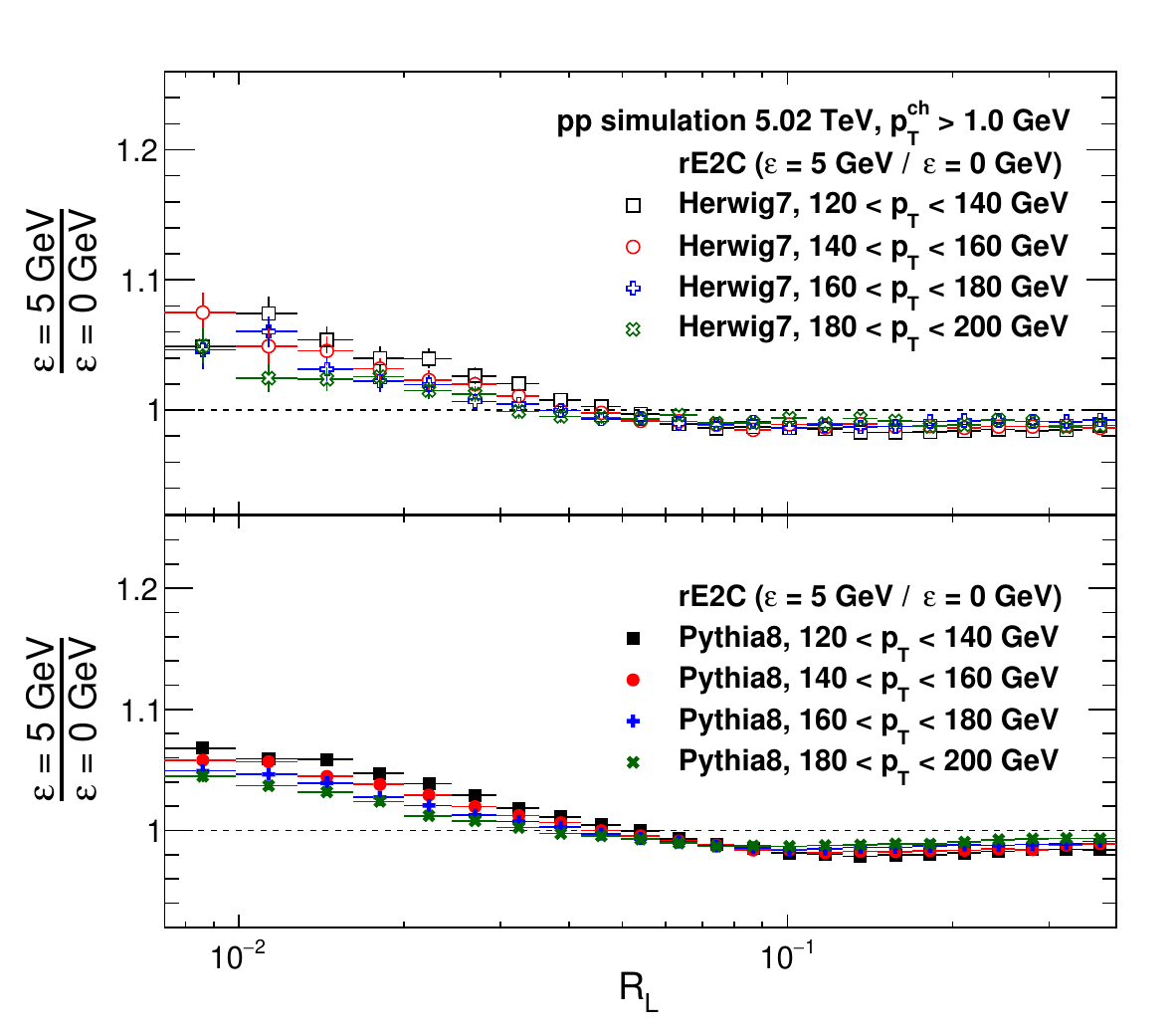}
    \includegraphics[width=0.49\linewidth]{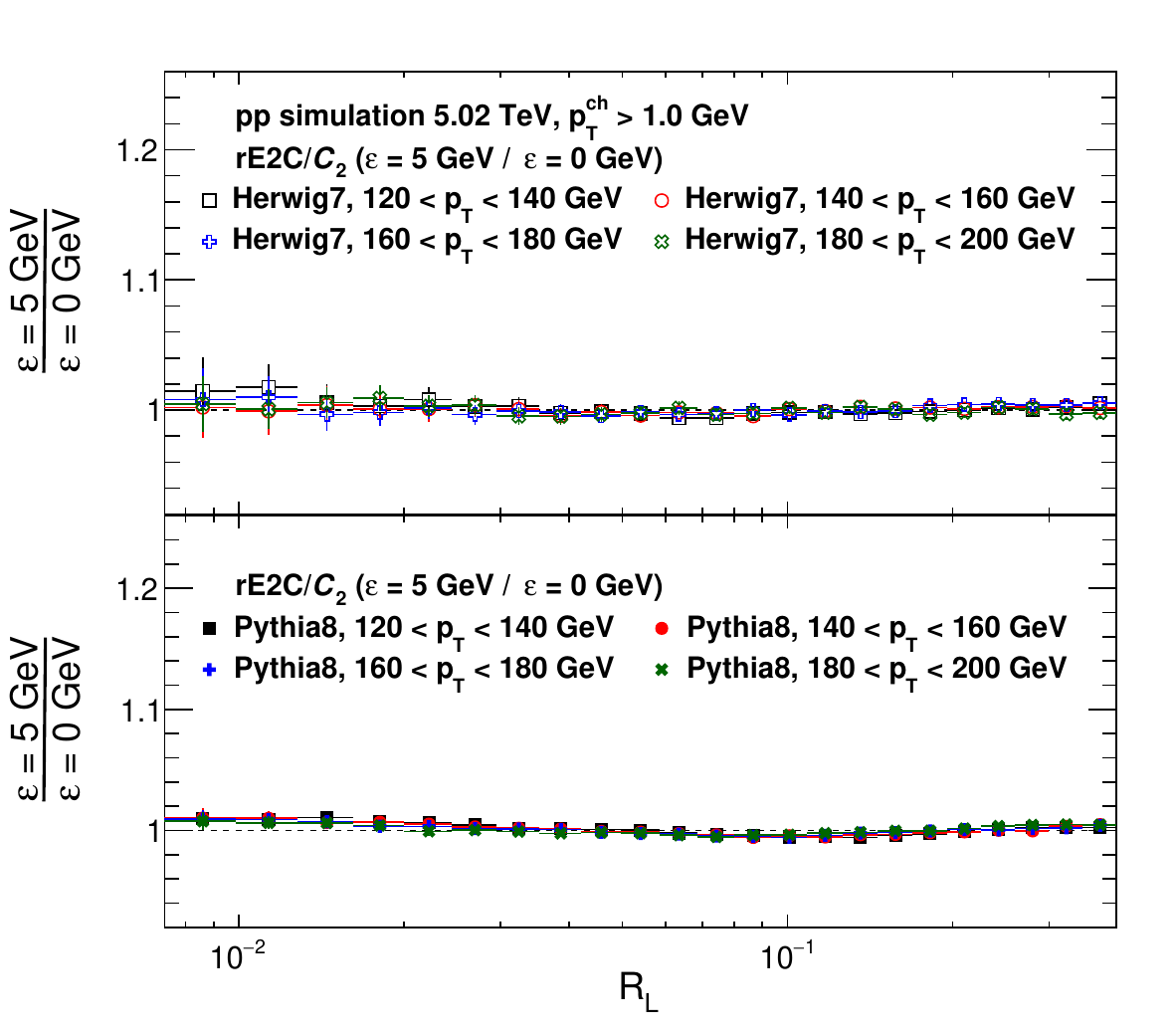}
    \caption{Left panel: Top (bottom) shows the E2C for inclusive \HERWIG (\PYTHIA) jets with $125 < p_{T} < 145~\text{GeV}$ (black), $145 < p_{T} < 165~\text{GeV}$ (red), $165 < p_{T} < 185~\text{GeV}$ (blue), and $185 < p_{T} < 205~\text{GeV}$ (green), each divided by the corresponding E2C for inclusive \HERWIG (\PYTHIA)  jets with $120 < p_{T} < 140~\text{GeV}$, $140 < p_{T} < 160~\text{GeV}$, $160 < p_{T} < 180~\text{GeV}$, and $180 < p_{T} < 200~\text{GeV}$, respectively. Right panel:  Top (bottom) shows the ratio E2C/$C_2$ for inclusive \HERWIG (\PYTHIA) jets with $125<p_T <145~$GeV (black), $145<p_T <165~$GeV (red), $165<p_T <185~$GeV (blue), and $185<p_T <205~$GeV (green),  each divided by the corresponding E2C for inclusive \HERWIG (\PYTHIA) jets with $120<p_T <140~$GeV,  $140<p_T <160~$GeV, $160<p_T <180~$GeV, $180<p_T <200~$GeV. This figure illustrates  the impact of  a $\varepsilon=5~\text{GeV}$ selection bias.}
    \label{fig:5GeVsupp}
\end{figure}

For brevity, we only include the ratio plots, corresponding to the bottom panels in Figs.~\ref{fig:5GeV} and \ref{fig:15GeV}. Fig.~\ref{fig:5GeVsupp} shows results for a small constant selection bias of $\varepsilon= 5$~GeV applied to anti-$k_{t}$ jets across several reconstructed $p_{T}$ $20~$GeV ranges from $120$ to $200$~GeV. The left panel displays the rE2C ratio defined  as the  E2C for these jets divided by that of the corresponding jets without selection bias, with the unbiasing function $C_{2}$ not applied. In contrast, the right panel presents the rE2C/$C_2$-ratio defined as the E2C/$C_2$ for the biased jets divided by the E2C of the corresponding jets without selection bias. The comparison between the two panels demonstrates that the inclusion of the unbiasing function $C_{2}$ effectively mitigates the selection bias in these E2C distributions.

\begin{figure}[ht]
    \centering
    \includegraphics[width=0.49\linewidth]{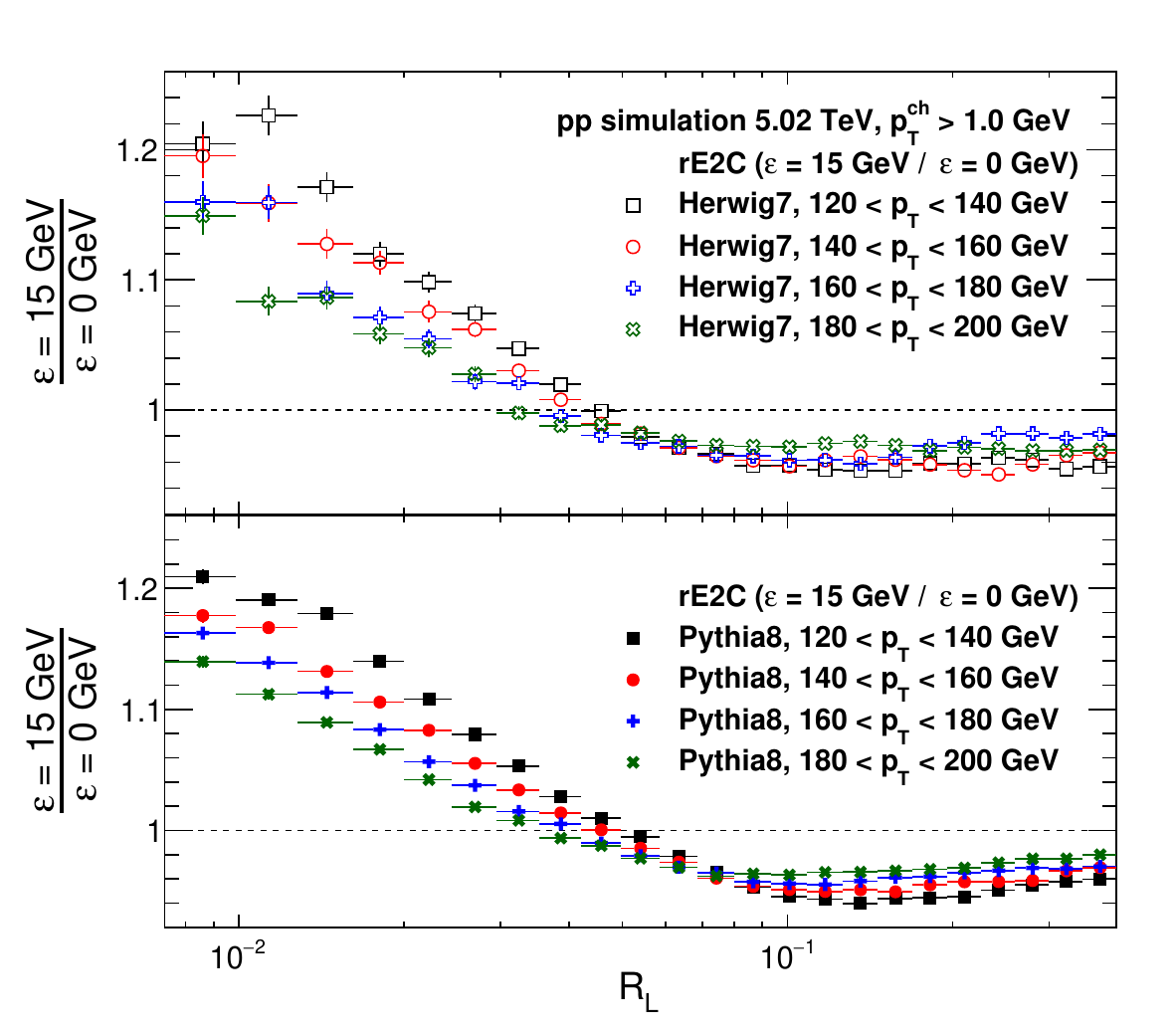}
    \includegraphics[width=0.49\linewidth]{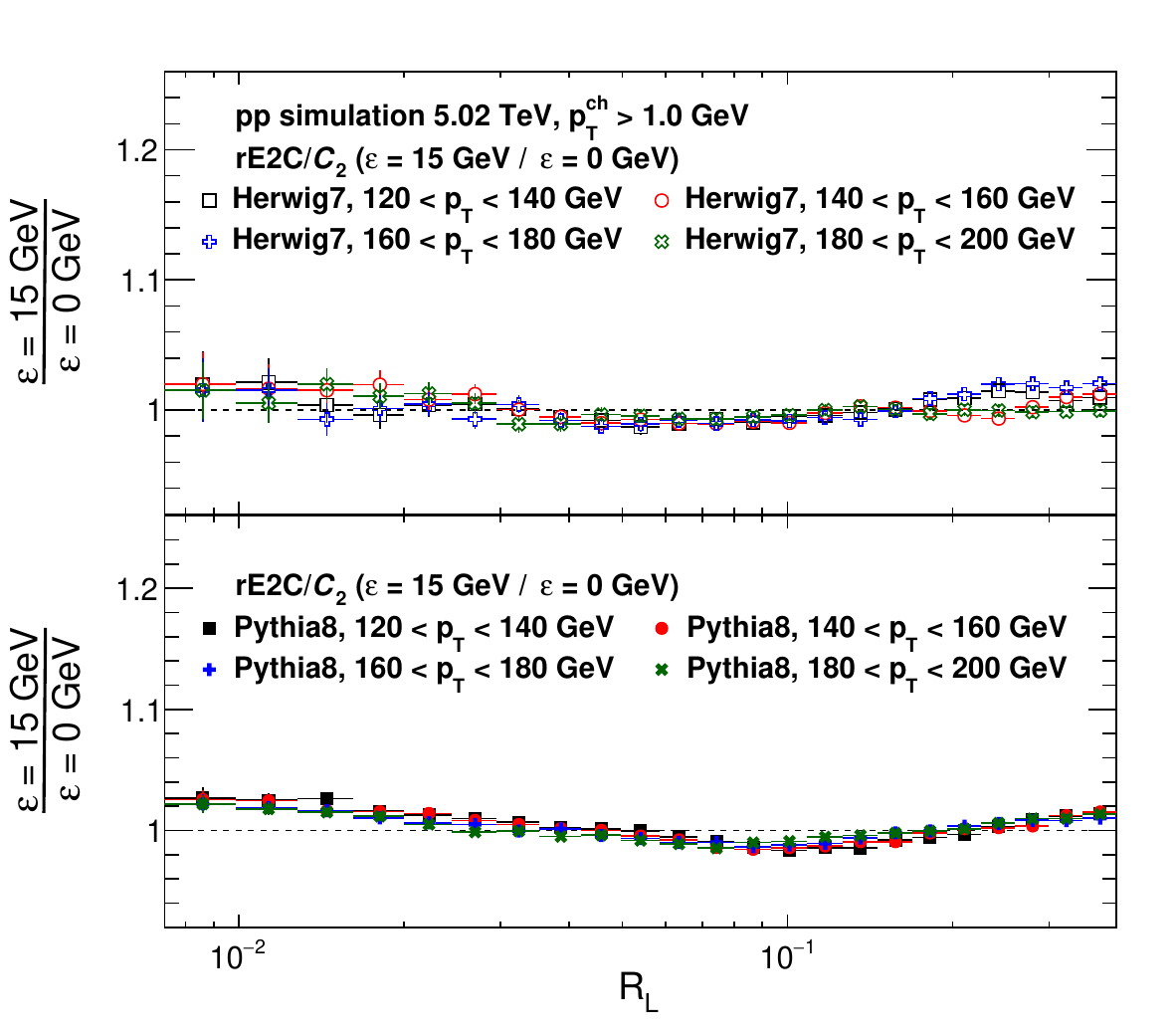}
    \caption{Left panel: Top (bottom) shows the E2C for inclusive \HERWIG (\PYTHIA) jets with $135 < p_{T} < 155~\text{GeV}$ (black), $155 < p_{T} < 175~\text{GeV}$ (red), $175 < p_{T} < 195~\text{GeV}$ (blue), and $195 < p_{T} < 215~\text{GeV}$ (green), each divided by the corresponding E2C for inclusive \HERWIG (\PYTHIA) jets with $120 < p_{T} < 140~\text{GeV}$, $140 < p_{T} < 160~\text{GeV}$, $160 < p_{T} < 180~\text{GeV}$, and $180 < p_{T} < 200~\text{GeV}$, respectively. Right panel:  Top (bottom) shows the ratio E2C/$C_2$ for inclusive \HERWIG (\PYTHIA) inclusive jets with $135<p_T <155~$GeV (black), $155<p_T <175~$GeV (red), $175<p_T <195~$GeV (blue), and $195<p_T <215~$GeV (green),  each divided by the corresponding E2C for inclusive \HERWIG (\PYTHIA) jets with $120<p_T <140~$GeV,  $140<p_T <160~$GeV, $160<p_T <180~$GeV, $180<p_T <200~$GeV.  This figure illustrates  the impact of  a $\varepsilon=15~\text{GeV}$ selection bias.}
    \label{fig:15GeVsupp}
\end{figure}

Fig.~\ref{fig:15GeVsupp} is analogous to Fig.~\ref{fig:5GeVsupp}, but for a larger constant selection bias of $\varepsilon= 15$~GeV. 
We observe that the unbiasing factor reduces the selection bias from a $\lesssim 20\%$ effect in the E2C  (left panel) to a $\lesssim 2\%$ effect in the E2C/$C_{2}$ distribution (right panel).

\end{document}